\documentclass[letter, a4paper]{IEEEtran}
\usepackage{amsmath,amsfonts}
\usepackage{algorithm}
\usepackage{array}
\usepackage{textcomp}
\usepackage{stfloats}
\usepackage{url}
\usepackage{verbatim}
\usepackage{graphicx}
\usepackage{cite}
\usepackage{float} 
\usepackage{authblk}
\usepackage{lipsum}
\usepackage[mathscr]{euscript}
\usepackage{framed}
\usepackage{algpseudocode}
\usepackage{bm}
\usepackage{stfloats}  
\usepackage{setspace}
\usepackage{balance}
\usepackage{amsthm} 

\ifCLASSOPTIONcompsoc
\usepackage[caption=false,font=normalsize,labelfont=sf,textfont=sf]{\part{title}subfig}
\else
\usepackage{subfigure}
\fi

\usepackage{graphicx,amsmath,amssymb,amsfonts}
\allowdisplaybreaks[3]
\usepackage{caption} 
\usepackage{hyperref}
\hypersetup{colorlinks=true}   
\usepackage{xcolor}
\usepackage[T1]{fontenc}   

\newtheorem{lemma}{\textbf{Lemma}}

\newtheorem{corollary}{\textbf{Corollary}}

\floatname{algorithm}{Algorithm}

\makeatletter
\renewcommand{\maketag@@@}[1]{\hbox{\m@th\normalsize\normalfont#1}}%
\makeatother

\newcommand{\E}{\mathbb{E}}

\renewcommand{\Re}{\operatorname{Re}}

\makeatletter

\newcommand{\Rmnum}[1]{\expandafter\@slowromancap\romannumeral #1@}

\makeatother

\begin{document}
	\captionsetup{font={small}}
	\bstctlcite{reference:BSTcontrol}
	
	\title{\fontsize{22 pt}{\baselineskip}\selectfont Joint Transmit and Receive Antenna Orientation Design for Secure MIMO Communications} 
	\author{
		\fontsize{10 pt}{\baselineskip}\selectfont Ailing~Zheng, Qingqing~Wu, Xingxiang~Peng, Qiaoyan~Peng, Ziyuan~Zheng, Yuxuan~Chen, Wen~Chen
		\vspace{-10 mm}
		\thanks{A. Zheng, Q. Wu, X. Peng, Z. Zheng, and W. Chen are with School of Information Science and Electronic Engineering, Shanghai Jiao Tong University, Shanghai 200240, China (e-mail: (ailing.zheng, qingqingwu, peng\_xingxiang, zhengziyuan2024, wenchen)@sjtu.edu.cn).}
		\thanks{Q. Peng is with School of Information Science and Electronic Engineering, Shanghai Jiao Tong University, Shanghai 200240, China, and also with the State Key Laboratory of Internet of Things for Smart City and the Department of Electrical and Computer Engineering, University of Macau, Macao SAR, China (email: qiaoyan.peng@connect.um.edu.mo).}
		\thanks{Y. Chen is with Huawei Device Company Ltd., Shenzhen 518129, China (email:  chenyuxuan56@huawei.com).}
		
	}
	\vspace{-10 mm}
	
	\maketitle
	\vspace{-5 mm}
	\begin{abstract}
		Physical layer security (PLS) is a promising paradigm for safeguarding 6G wireless networks by exploiting the inherent characteristics of wireless channels. However, the efficiency of conventional PLS is often limited by fixed orientation antennas. 
		This paper investigates a rotatable antenna (RA)-aided secure multiple-input multiple-output (MIMO) communication system, where both the transmitter and the  receiver are equipped with RAs in the presence of an eavesdropper.  
		By dynamically optimizing the orientations of RAs, we can proactively reshape the effective MIMO channels to enhance legitimate transmission while simultaneously suppressing information leakage to the eavesdropper.
		We formulate a secrecy rate maximization problem by jointly optimizing the transmit beamforming, artificial noise (AN) covariance matrix, and the transmit/receive RA orientations, subject to the transmit power budget and antenna orientation constraints.
		To tackle the resulting highly coupled and non-convex problem, we first study a simplified single-input single-output (SISO) case to reveal the structure of the optimal RA orientation. For the general MIMO case, we develop an alternating optimization algorithm by reformulating the original problem through the minimum mean-square error framework. In particular, the transmit beamforming and AN covariance matrix are derived in semi-closed forms, while the RA orientations are updated via the Riemannian Frank-Wolfe method. The proposed design is further extended to the multi-receiver secure transmission scenario. Simulation results show that the proposed scheme converges rapidly and achieves significant secrecy rate gains over the conventional fixed-orientation scheme.
	\end{abstract}
	
	\begin{IEEEkeywords}
		Rotatable antenna (RA), physical layer security (PLS), artificial noise (AN), antenna orientation optimization.
	\end{IEEEkeywords}
	
	\vspace{-3mm}
	\section{Introduction}
	\vspace{-1mm}
	With the rapid advancement of global information and communication technology, future wireless networks are expected to support intelligent connectivity for a dramatically growing number of devices and users, leading to the generation and transmission of enormous amounts of data \cite{6Gvisions2023}. Consequently, due to the broadcast nature of wireless channels, confidential information is inherently vulnerable to unauthorized interception. Therefore, secure communication has become increasingly important in sixth generation (6G) communication networks \cite{6Gsecure2021}. Although conventional cryptographic techniques can provide security at upper layers, they usually rely on high computational complexity and sophisticated key management, which may become less effective in highly dynamic and large-scale wireless systems \cite{stallings2010cryptography}. As a complementary approach, physical layer security (PLS) has attracted significant attention by exploiting the physical characteristics of wireless channels to enhance secure transmission~\cite{Shiu2011}. 
	
	Over the last few decades, various PLS techniques have been developed to improve secure communication performance.
	Specifically, multi-antenna technology can be used to design effective beamforming toward legitimate receivers while suppressing signal energy in unintended directions \cite{Khisti2010,Lisecure2017,Liusecure2014}. Artificial noise (AN)-based jamming provides another solution by interfering with the eavesdroppers to increase the secrecy rate (SR), where the specific interfering signals are generated according to channel characteristics to reduce the wiretap channel capacity without affecting the legitimate channel capacity \cite{NiuPLS2026}. Compared to imposing the AN in the null space of legitimate channels, which offers restricted optimization space \cite{TsaiAN2014}, the joint design of AN and the transmit beamforming has gained more attention. For instance, the authors of \cite{LinAN2013} proposed a generalized AN design, allowing the injection of AN to the legitimate channel, which verified the significant advantage of the generalized AN design over the conventional AN approach. 
	
	However, existing PLS techniques mainly rely on transmit-side signal processing. By properly shaping the transmitted signal, these methods aim to enhance the quality of the legitimate link while suppressing the information leakage to eavesdroppers. Unfortunately, these technologies are inherently constrained by the conventional antenna architecture with fixed positions and orientations. As a result, the available degrees of freedom (DoFs) for secrecy enhancement are mainly limited to the digital domain, while the spatial directivity of antennas is not fully exploited. In challenging propagation environments, especially when the legitimate and eavesdropping channels are highly correlated, such conventional approaches may suffer from limited secrecy performance.
	To break these bottlenecks, movable antenna (MA) has been proposed, where the antenna positions can be flexibly adjusted to reshape the channel characteristics \cite{Zhu2024MA}. Particularly, to support secure communication, antennas are dynamically repositioned to optimize channel conditions for legitimate users while simultaneously ensuring that eavesdroppers are relegated to regions of deep fading \cite{MasecureMA2026,Hu2024,Tang2025PLS}. Several initial works have demonstrated the potential of MAs in improving the secrecy communication rate by allowing one-dimensional (1D) antenna movement \cite{Hu2024MA} or three-dimensional (3D) flexibility \cite{Ma2024MA} in antenna positioning. 
	
	However, despite its promising performance, the application of MA in fast-varying channels is impeded by its limited response time and movement speed. Additionally, conventional MAs are restricted to positional adjustments alone, as the antenna orientation remains fixed. This limitation prevents them from fully adapting to dynamic channel conditions. To address these drawbacks, the concept of six-dimensional (6D) MA architecture was proposed in \cite{Shao2025-1} and \cite{Shao2025}, where both the position and rotation of each antenna surface can be optimized to track varying channel conditions. The superiority of 6DMA has been demonstrated in improving network capacity \cite{Shao2025-2}, securing communication \cite{Qian2025secure6DMA}, and spatial multiplexing \cite{Shi2025MA}. As a lightweight implementation of 6DMA, rotatable antennas (RAs) have emerged as a promising hardware architecture to reduce the hardware complexity and support compact design \cite{ZhengRA2026}. By dynamically adjusting the 3D boresight direction of each antenna, RA-enabled systems can actively reshape effective channel gains in the angular domain, thereby collaboratively enhancing the overall array gain \cite{Peng2025,Zheng202520,Zheng2026}. Compared with conventional fixed-orientation antennas, RAs provide additional spatial DoFs by adaptively shaping the radiation direction, thereby concentrating energy toward desired angular regions or spatial locations and enhancing the effective array gain for intended users.
	Motivated by these advantages, several works have investigated RA-enabled wireless communication systems in integrated sensing and communication \cite{Zhou2025,Zheng2025MA-4}, secure communication \cite{Jiang2025RA,Liang2025}, and unmanned aerial vehicle systems~\cite{Zhang2025RA}.

	By intelligently reconfiguring antenna orientations, RAs can strengthen desired signal transmission toward legitimate users while reducing signal leakage toward unintended directions, thereby improving communication security by enlarging the channel quality difference between legitimate users and eavesdroppers. Recently, several preliminary works have begun to explore RA-assisted secure communications.
	In particular, the authors in \cite{Jiang2025RA} investigated a single-input single-output (SISO) system involving one legitimate receiver and one eavesdropper, where a closed-form expression for the optimal deflection angle was derived to maximize the secrecy capacity. The RA-aided secure multiple-input single-output (MISO) system was studied in \cite{Liang2025}, where one legitimate receiver and multiple eavesdroppers were considered. 
	Meanwhile, with the continuous evolution of antenna technology, wireless devices equipped with multiple antennas are expected to play an increasingly important role in future networks \cite{Jha2023}. In view of this development trend, it is of great interest to extend RA-assisted secure communication design to more general multi-antenna scenarios. Therefore, we investigate a more general RA-aided secure multiple-input multiple-output (MIMO) communication system with the assistance of AN in this paper.

	Despite the promising benefits of RA-aided secure MIMO systems, several critical technical challenges remain to be addressed. First, the introduction of RAs into MIMO architectures significantly complicates channel modeling, as the spatial directivity becomes a dynamic function of the antenna's 3D orientation under multipath propagation. Second, the resulting SR maximization problem involves the joint optimization of transmit beamforming, AN covariance matrix, and RA orientations. These variables are highly coupled within the objective function, leading to a complex non-convex optimization problem. Furthermore, practical hardware limitations, such as constraints on the rotation range, impose additional non-convex boundaries. Developing an efficient, low-complexity algorithm to decouple these variables and find a near-optimal solution while ensuring robust secrecy performance constitutes a major theoretical and practical hurdle.
	
	In this paper, we study an RA-aided secure MIMO communication system, where both the transmitter and the legitimate receiver are equipped with multiple RAs, while an eavesdropper attempts to intercept the transmitted message. 
	The main contributions of this paper are summarized as follows:
	\begin{itemize}
		\item We establish a general RA-aided secure MIMO communication model under multipath propagation, where both the transmitter and the receiver are equipped with RAs. Then, we formulate a SR maximization problem by jointly optimizing the transmit beamforming, AN covariance matrix, and RA orientations, subject to the transmit power budget and antenna orientation constraints.
		\item 
		We first study a simplified SISO line-of-sight (LoS)-dominant case to uncover the
		secrecy mechanism of RA rotation. Without zenith angle constraints, the
		transmit orientation can be characterized in the plane spanned by the
		legitimate and eavesdropping directions. This leads to a simple
		leakage-nulling rotation rule that can force the eavesdropping gain to zero
		while preserving a nonzero legitimate-link gain when the nulling angle is
		feasible, thereby converting an unfavorable fixed-orientation SISO wiretap
		channel into a positive-secrecy channel. Under practical zenith angle
		constraints, a 1D search is developed to characterize the
		tradeoff between legitimate-link enhancement and leakage suppression.	
		\item For the general MIMO case, we propose an efficient alternating optimization (AO) algorithm to tackle the formulated non-convex problem. Specifically, we first transform the original problem into a more tractable form by using the minimum mean-square error (MMSE) framework. Then, the optimal transmit beamforming and AN covariance matrix are obtained by applying the Lagrangian multiplier method in semi-closed forms. The Riemannian Frank-Wolfe method is exploited to obtain the transmit/receive RA orientations.
		We further extend the proposed framework to the multi-receiver multicast secure transmission scenario. 
		To handle the resulting worst-receiver SR, we derive a tractable lower bound of the multicast SR and develop an AO-based algorithm to maximize this lower-bound objective.
		\item Simulation results demonstrate the convergence of the proposed algorithm. The proposed algorithm outperforms the isotropic-antenna baseline by $61\%$. Moreover, more data streams can significantly improve the SR in the high-transmit-power region, where the scheme with $4$ streams achieves a $28\%$ gain over that with $2$ streams. Additionally, the proposed algorithm attains
		a 20\% performance gain compared to the discrete-orientation baseline.
	\end{itemize}
	
	The remainder of this paper is organized as follows. Section~\ref{System Model} introduces the system model and formulates the SR maximization problem. Section~\ref{Single antenna case} studies the SISO case and provides useful analytical insights. Section~\ref{Proposed Solution} presents the proposed AO-based solution for the general MIMO scenario. Section~\ref{Extension to The Multi-Receivers Case} extends the framework to the multi-receiver case. Section~\ref{Simulation Results} provides simulation results, and Section~\ref{Conclusion} concludes this paper.

	\vspace{-4mm}
	\section{System Model}   \label{System Model}
	\vspace{-2mm}
	\begin{figure}[t]
		\centering
		\includegraphics[width=0.42 \textwidth]{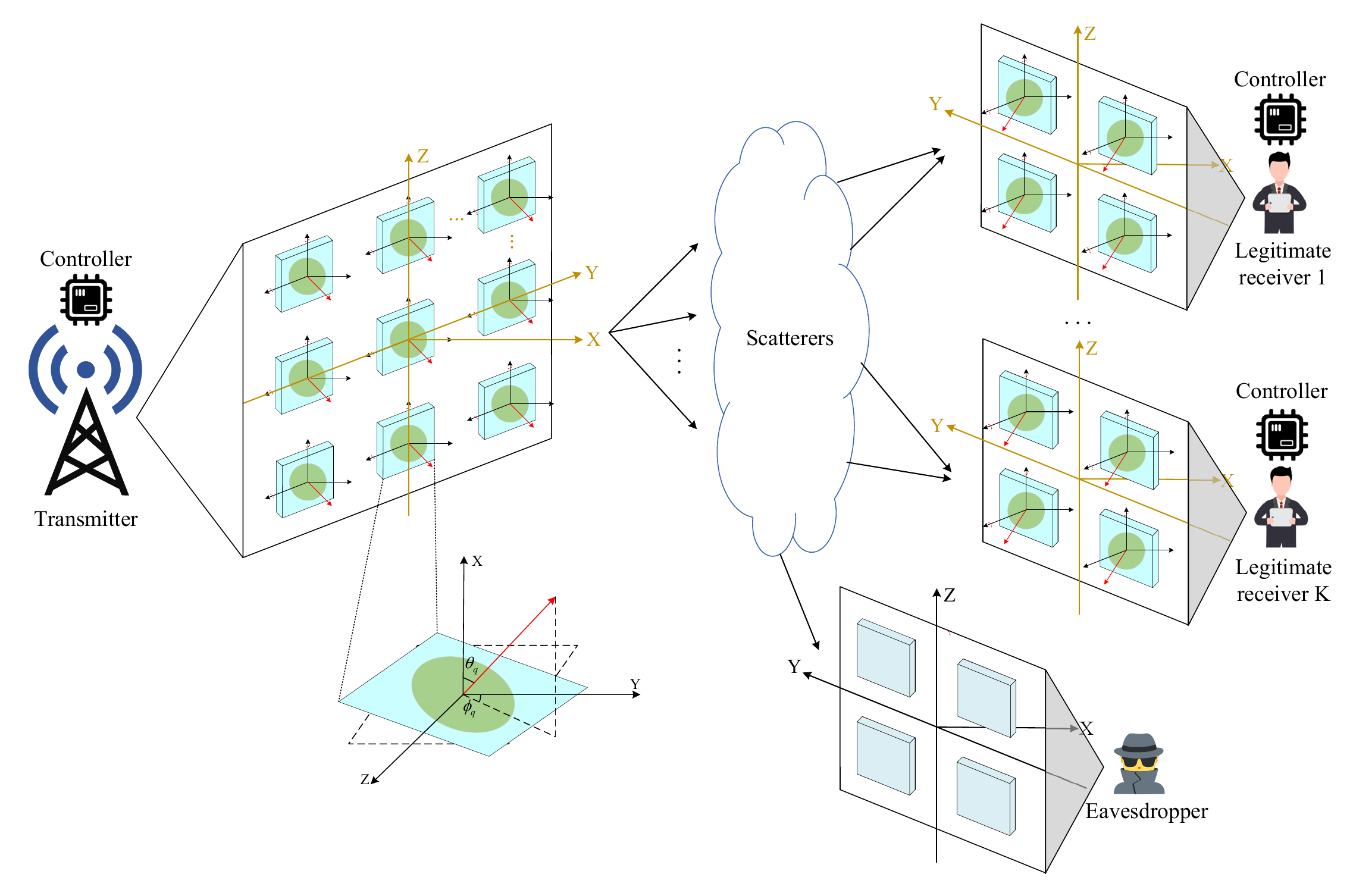}
		\caption{Illustration of the RA-aided secure MIMO communication architecture, where both the transmitter and legitimate receivers are equipped with RAs, while the eavesdropper is equipped with fixed isotropic antennas.}
		\label{Fig1}
		\vspace{-8mm}
	\end{figure}
	As shown in Fig.~\ref{Fig1}, we consider a general RA-aided secure communication architecture.
	For ease of exposition, we first focus on the fundamental wiretap scenario with one legitimate receiver and one eavesdropper. 
	The resulting single-receiver model provides the basis for the proposed channel modeling and optimization framework, which is further extended to the multi-receiver secure transmission scenario in Section~\ref{Extension to The Multi-Receivers Case}.
	Specifically, the transmitter is equipped with $N$ transmit RAs, the legitimate receiver consists of $M$ receive RAs, and the eavesdropper is equipped with $Q$ fixed isotropic antennas.
	The uniform planar array (UPA) at the transmitter (receiver or eavesdropper) is located in its local $x_t-y_t$ ($x_{r}-y_{r}$ or $x_{e}-y_{e}$) plane, satisfying $N=N_x \times N_y$, $M=M_{x} \times M_{y}$, and $Q=Q_{x} \times Q_{y}$, respectively. Their global positions are $\mathbf{t}_0=(x_t,y_t,z_t)$, $\mathbf{r}_0=(x_{r},y_{r},z_{r})$, and $\mathbf{e}_0=(x_{e},y_{e},z_{e})$, respectively. Denote $\mathcal{N}=\{1,2,\ldots, N\}$, $\mathcal{M}=\{1,2,\ldots, M\}$, and $\mathcal{Q}=\{1,2,\ldots, Q\}$ as the sets of antennas at the transmitter, receiver, and eavesdropper, respectively.
	The local positions of the $n$-th transmit RA and the $m$-th receive RA can be characterized by $\mathbf{t}_{n}^l = [x_{t,n}, y_{t,n}, z_{t,n}]^\mathrm{T}$ and $\mathbf{r}_{m}^l = [x_{r,m}, y_{r,m}, z_{r,m}]^\mathrm{T}$, respectively, which are
	\setlength\abovedisplayskip{-1pt}
	\setlength\belowdisplayskip{0pt}
	\begin{align}
		&\!\!\!\! x_{t,n} \!=\! (n_x \!-\! \frac{N_x \!+\!1}{2})\Delta, y_{t,n} \!=\!(n_y-\frac{N_y \!+\! 1}{2})\Delta, 
		z_{t,n} \!=\!0, \nonumber \\
		&\!\!\!\! x_{r,m} \!=\! (m_{x} \!-\!\frac{M_{x} \!+ \!1}{2})\Delta, y_{r,m} \!=\!(m_{y} \!-\! \frac{M_{y} \!+\! 1}{2})\Delta, 
		z_{r,m} \!= \!0,  \nonumber 
	\end{align}
	where $n=n_x+(n_y-1)N_x$, $m = m_{x}+(m_{y}-1)M_x$, and $\Delta$ is the inter-element spacing. Similarly, the local position of the $q$-th antenna at the eavesdropper is $\mathbf{e}_{q}^l = [x_{e,q}, y_{e,q}, z_{e,q}]^\mathrm{T}$.
	
	Let $\mathbf{R}_t$, $\mathbf{R}_{r}$, and $\mathbf{R}_{e}$ denote the rotation matrix from the local coordinate system to the global coordinate system for the transmitter, receiver, and eavesdropper, respectively. Define $\mathbf{t}_n$, $\mathbf{r}_{m}$, and $\mathbf{e}_{q}$ as the global position of transmit RA $n$, the receive RA $m$, and the eavesdropper antenna $q$, respectively. Then, we have $	\mathbf{t}_n = \mathbf{t}_0+\mathbf{R}_t\mathbf{t}_n^l,\mathbf{r}_{m} = \mathbf{r}_0+\mathbf{R}_{r}\mathbf{r}_{m}^l$, and $\mathbf{e}_{q} = \mathbf{e}_0+\mathbf{R}_{e}\mathbf{e}_{q}^l$.
	We assume that the position and posture of each UPA remain fixed. Thus, the rotation matrices $\mathbf{R}_t$, $\mathbf{R}_{r}$, and $\mathbf{R}_{e}$ are predetermined during deployment. However, each antenna element on the UPA can independently adjust its orientation through electronic or mechanical means \cite{Peng2025}.
	Denote the 3D boresight direction of transmit RA $n$ in the local coordinate system as
	$\mathbf{f}_{t,n} = [f_{t,n}^x,f_{t,n}^y,f_{t,n}^z]^T$,
	where $f_{t,n}^x$, $f_{t,n}^y$, and $f_{t,n}^z$ are the projections of transmit RA $n$'s orientation on the $x$-, $y$-, and $z$-axes, respectively. Let $\theta_{z,n} $ denote the RA $n$'s zenith angle (i.e., the angle between the boresight direction of RA $n$ and the $z$-axis) and $\theta_{a,n}$ represent its azimuth angle in the local coordinate system. Thus, we have $f_{t,n}^x = \sin\theta_{z,n}\cos\theta_{a,n}$, $f_{t,n}^y = \sin\theta_{z,n}\sin\theta_{a,n}$, and $f_{t,n}^z = \cos\theta_{z,n}$. Furthermore, we have $\Vert \mathbf{f}_{t,n} \Vert_2 = 1$ due to normalization. To account for practical rotational constraint and mitigate antenna coupling between any two RAs, the zenith angle of each RA should be confined to a specific range
	$	0 \le \theta_{z,n} \le \theta_{\max}$,
	where $\theta_{\max} \in [0, \pi/2]$ is the maximum zenith angle that each RA is allowed to adjust. This angular constraint on $\theta_{z,n}$ is equivalent to the following constraint on the boresight vector
	\begin{align}
		\label{theta_max}
		\cos(\theta_{\max}) \le \mathbf{f}_{t,n}^T\mathbf{e}_z \le 1,
	\end{align}
	where $\mathbf{e}_z = [0,0,1]^T$ is the unit vector along the $z$-axis.
	Similarly, the orientation of receive RA $m$ is given by $\mathbf{f}_{r,m}$.
	The corresponding boresight vector in the global coordinate system is expressed as 
	$\tilde{\mathbf{f}}_{t,n} = \mathbf{R}_{t}\mathbf{f}_{t,n}$ and $\tilde{\mathbf{f}}_{r,m} = \mathbf{R}_{r}\mathbf{f}_{r,m}$.
	Let $\mathbf{f}_{t} = [\mathbf{f}_{t,1},\mathbf{f}_{t,2},\ldots,\mathbf{f}_{t,N}] \in \mathcal{R}^{3 \times N}$ and $\mathbf{f}_{r} = [\mathbf{f}_{r,1},\mathbf{f}_{r,2},\ldots,\mathbf{f}_{r,M}] \in \mathcal{R}^{3 \times M}$ as the collections of the orientations of $N$ transmit RAs and $M$ receive RAs, respectively. 
	
	The effective antenna gain for each RA depends on the signal arrival/departure angle and antenna directional gain pattern, where $\epsilon$ is the angular offset between the signal direction and the antenna's main-lobe boresight. 
	The generic directional gain pattern for each RA is as follows:
	\begin{align}
		\label{antennaGain}
		G(\epsilon) = 
		\begin{cases}
			G_0\cos^{2p}(\epsilon), & \epsilon \in [0, \frac{\pi}{2}], \\
			0, & \text{otherwise},
		\end{cases}
	\end{align}
	where $p \ge 0$ determines the antenna directivity, and $G_0$ is the maximum gain in the boresight direction with $G_0=2(2p+1)$ to satisfy the law of power conservation.
	
	We adopt a narrow-band geometric frequency-flat channel model~\cite{ Chen2022IRS}. The perfect channel state information (CSI) is assumed to reveal the fundamental performance of the proposed RA-aided secure MIMO architecture~\cite{Xiong2025RA}. The LoS link power gain between the transmit RA $n$ to the receive RA $m$ is
	\begin{align}
		& G_{m,n}^{\mathrm{LoS}}(\mathbf{f}_{t,n},\mathbf{f}_{r,m}) \!=\!\beta_0r_{m,n}^{-2}G(\epsilon_{m,n})G(\epsilon_{n,m}) \nonumber \\ 
		&\!\!\! =\!\beta_0r_{m,n}^{-2}G_0^2\left[\!\frac{\tilde{\mathbf{f}}_{t,n}^\mathrm{T}(\mathbf{r}_{m} \!-\! {\mathbf{t}}_{n})}{r_{m,n}}\!\right]_{+}^{2p} \left[\!\frac{\tilde{\mathbf{f}}_{r,m}^\mathrm{T}(\mathbf{t}_n \! -\! {\mathbf{r}}_{m})}{r_{m,n}}\!\right]_{+}^{2p}, \nonumber
	\end{align}
	where $\beta_0 = (\frac{\lambda}{4\pi })^2$ is the free-space reference gain constant, $\lambda$ denotes the wavelength, and $r_{m,n} = \Vert {\mathbf{t}_n - \mathbf{r}}_{m}\Vert_2$ is the distance between the transmit RA $n$ and the receive RA $m$. Here, $\cos(\epsilon_{m,n}) =\frac{\tilde{\mathbf{f}}_{t,n}^\mathrm{T}({\mathbf{r}}_{m}-\mathbf{t}_n)}{r_{m,n}}$ represents the cosine of the angle between the transmit RA $n$'s orientation $\tilde{\mathbf{f}}_{t,n}$ and the LoS direction to the receive RA $m$. Similarly, $\cos(\epsilon_{n,m}) =\frac{\tilde{\mathbf{f}}_{r,m}^\mathrm{T}(\mathbf{t}_n-\mathbf{r}_{m})}{r_{m,n}}$ denotes the cosine of the angle between the receive RA $m$'s orientation $\tilde{\mathbf{f}}_{r,m}$ and the LoS direction to the transmit RA $n$. The related channel is expressed as \cite{Zheng2026}
	\begin{align}
		h_{m,n}^{\mathrm{LoS}}(\mathbf{f}_{t,n},\mathbf{f}_{r,m})&= \sqrt{G_{m,n}^{\mathrm{LoS}}(\mathbf{f}_{t,n},\mathbf{f}_{r,m})}e^{-j\frac{2\pi}{\lambda}r_{m,n}} \nonumber \\
		& = \sqrt{\beta_0}r_{m,n}^{-1}G_0{f}_{t,m,n}{f}_{r,m,n}e^{-j\frac{2\pi}{\lambda}r_{m,n}}, \nonumber \\
		& = a_{m,n}{f}_{t,m,n}{f}_{r,m,n},
	\end{align}
	where $a_{m,n}=\frac{\sqrt{\beta_0}G_0}{r_{m,n}}e^{-j\frac{2\pi}{\lambda}r_{m,n}}$ and
	\begin{align}
		&\!\!\!\!\! {{f}}_{t,m,n} \!\!=\!\! \left[\frac{\tilde{\mathbf{f}}_{t,n}^\mathrm{T}(\mathbf{r}_{m} \!- \!{\mathbf{t}}_{n})}{r_{m,n}}\right]_{+}^{p},  {{f}}_{r,m,n} \!\!=\!\! \left[\frac{\tilde{\mathbf{f}}_{r,m}^\mathrm{T}(\mathbf{t}_n \! - \!{\mathbf{r}}_{m})}{r_{m,n}}\right]_{+}^{p}.
	\end{align}
	The channel coefficient captures both the directional antenna gain and the propagation-included phase shift.
	
	We further consider a scattering environment with $D$ spatially distributed clusters, located at $\{\mathbf{s}_d \in \mathbb{R}^3\}^D_{d=1}$ in the global coordinate system. The non-LoS (NLoS) link power gain between transmit RA $n$ and cluster $d$ is
	\begin{align}
		G_{n,d}^{\mathrm{NLoS}}(\mathbf{f}_{t,n}) & =\beta_0r_{n,d}^{-2}G_0\left[\frac{\tilde{\mathbf{f}}_{t,n}^\mathrm{T}(\mathbf{s}_d-{\mathbf{t}}_{n})}{r_{n,d}}\right]_{+}^{2p},
	\end{align}
	where $r_{n,d} =\Vert \mathbf{s}_d-{\mathbf{t}}_{n} \Vert_2$ is the antenna-to-cluster distance, and $\cos(\epsilon_{n,d})= \frac{\tilde{\mathbf{f}}_{t,n}^\mathrm{T}(\mathbf{s}_d-{\mathbf{t}}_{n})}{r_{n,d}}$ denotes the cosine of the angle between the boresight and the direction to cluster $d$. Similarly, the NLoS link power gain between cluster $d$ and the receive RA $m$ is
	\begin{align}
		G_{d,m}^{\mathrm{NLoS}}(\mathbf{f}_{r,m}) & =\beta_0r_{d,m}^{-2}G_0\left[\frac{\tilde{\mathbf{f}}_{r,m}^\mathrm{T}(\mathbf{s}_d-{\mathbf{r}}_{m})}{r_{d,m}}\right]_{+}^{2p}.
	\end{align}
	Considering a bi-static scattering model \cite{Peng2025}, the NLoS channel coefficient from the transmit RA $n$ to the receive RA $m$ is given by
	\begin{align}
		&\!\! h_{m,n}^{\mathrm{NLoS}}(\mathbf{f}_{r,m},\mathbf{f}_{t,n}) \!=\! \nonumber \\
		&\!\!\sum \nolimits_{d=1}^D \!\! \sqrt{\frac{\sigma_dG_{n,d}^{\mathrm{NLoS}}(\mathbf{f}_{t,n})G_{d,m}^{\mathrm{NLoS}}(\mathbf{f}_{r,m})}{4\pi }} e^{-j\frac{2\pi}{\lambda}(r_{n,d}+r_{d,m})+j\chi_d}, \nonumber \\
		&\!\!= \!\!\sum \nolimits_{d=1}^D \!\! b_{m,n,d} \!\left[\!\frac{\tilde{\mathbf{f}}_{t,n}^\mathrm{T}(\mathbf{s}_d \!-\!{\mathbf{t}}_{n})}{r_{n,d}} \!\right]_{+}^{p}\left[\!\frac{\tilde{\mathbf{f}}_{r,m}^\mathrm{T}(\mathbf{s}_d \!-\!{\mathbf{r}}_{m})}{r_{d,m}}\!\right]_{+}^{p}, \nonumber\\
		&\!\! =\!\! \sum \nolimits_{d=1}^D b_{m,n,d}{f}_{t,n,d}{{f}}_{r,d,m},
	\end{align}
	where $b_{m,n,d} \!=\! \sqrt{\frac{\sigma_d}{4\pi}} \beta_0r_{n,d}^{-1}r_{d,m}^{-1}G_0e^{-j\frac{2\pi}{\lambda}(r_{n,d}+r_{d,m})+j\chi_d}$, with
	\begin{align}
		&\!\!\!\!\! {f}_{t,n,d} \!\!=\!\! \left[\frac{\tilde{\mathbf{f}}_{t,n}^\mathrm{T}(\mathbf{s}_d \!- \!{\mathbf{t}}_{n})}{r_{n,d}}\right]_{+}^{p}, 
	{{f}}_{r,d,m} \!\!=\!\! \left[\frac{\tilde{\mathbf{f}}_{r,m}^\mathrm{T}(\mathbf{s}_d \!- \!{\mathbf{r}}_{m})}{r_{d,m}}\right]_{+}^{p}.
	\end{align}
	The term $r_{d,m} = \Vert \mathbf{s}_d - \mathbf{r}_{m}\Vert_2$ denotes the cluster-to-receive RA $m$ distance, $\sigma_d$ denotes the radar cross section of cluster $d$, and $\chi_d$ is a random phase distributed over $[0, 2\pi)$. Then, we have
	\begin{align}
		h_{m,n} = h_{m,n}^{\mathrm{LoS}}(\mathbf{f}_{r,m},\mathbf{f}_{t,n})+h_{m,n}^{\mathrm{NLoS}}(\mathbf{f}_{r,m},\mathbf{f}_{t,n}). 
	\end{align}
	The overall multipath channel between the transmitter and the receiver is given by
	\begin{align}
		\label{channel_h}
		&\!\!\! \mathbf{H}(\mathbf{f}_{t},\mathbf{f}_{r}) \!=\! [\hat{\mathbf{h}}_1(\mathbf{f}_{t,1}), \hat{\mathbf{h}}_2(\mathbf{f}_{t,2}), \ldots, \hat{\mathbf{h}}_N(\mathbf{f}_{t,N})]\in \mathbb C^{M\times N},
	\end{align}
	where $\hat{\mathbf{h}}_n(\mathbf{f}_{t,n}) \!=\! [h_{1,n}(\mathbf{f}_{r,1}), h_{2,n}(\mathbf{f}_{r,2}), \ldots, h_{M,n}(\mathbf{f}_{r,M})]^\mathrm{T}$.
	
	Since the eavesdropper is equipped with fixed isotropic antennas, we construct the related channel as follows.
	The LoS link power gain between the transmit RA $n$ and the eavesdropper's antenna $q$ is given by
	\begin{align}
		G_{e,q,n}^{\mathrm{LoS}}(\mathbf{f}_{t,n}) &=\beta_0r_{e,q,n}^{-2}G(\epsilon_{e,q,n}) \nonumber \\ & ={\beta_0}r_{e,q,n}^{-2}G_0\left[\frac{\tilde{\mathbf{f}}_{t,n}^\mathrm{T}(\mathbf{e}_{q}-{\mathbf{t}}_{n})}{r_{e,q,n}}\right]_{+}^{2p},
	\end{align}
	where $r_{e,q,n} = \Vert {\mathbf{e}_{q} -\mathbf{t}_n}\Vert_2$ is the distance between the transmit RA $n$ and the eavesdropper's antenna $q$. The related channel is expressed as
	\begin{align}
		\!\!\!\! h_{e,q,n}^{\mathrm{LoS}}(\mathbf{f}_{t,n})&\!\!=\!\! \sqrt{G_{e,q,n}^{\mathrm{LoS}}(\mathbf{f}_{t,n})}e^{-j\frac{2\pi}{\lambda}r_{e,q,n}} \nonumber \\
		& \!\!=\!\! \sqrt{\beta_0G_0}r_{e,q,n}^{-1}f_{e,q,n}e^{-j\frac{2\pi}{\lambda}r_{e,q,n}}
		 \!\!=\!\! a_{e,q,n}f_{e,q,n},
	\end{align}
	where $a_{e,q,n}=\sqrt{\beta_0G_0}r_{e,q,n}^{-1}e^{-j\frac{2\pi}{\lambda}r_{e,q,n}}$ and $f_{e,q,n} = \left[\frac{\tilde{\mathbf{f}}_{t,n}^\mathrm{T}(\mathbf{e}_{q}-{\mathbf{t}}_{n})}{r_{e,q,n}}\right]_{+}^{p}$.
	Following the same principle, the NLoS link power gain between the transmit RA $n$ and the cluster $d$ is $G_{n,d}^{\mathrm{NLoS}}(\mathbf{f}_{t,n})$.
	The NLoS channel coefficient from the transmit RA $n$ to the eavesdropper's antenna $q$ is given by
	\begin{align}
		& h_{e,q,n}^{\mathrm{NLoS}}(\mathbf{f}_{t,n})  \!\!=\!\!\sum \nolimits_{d=1}^D \!\! \sqrt{\frac{\sigma_dG_{n,d}^{\mathrm{NLoS}}(\mathbf{f}_{t,n})}{4\pi r_{e,d,q}^2  }} e^{-j\frac{2\pi}{\lambda}(r_{n,d}+r_{e,d,q})\!+\! j\chi_d}, \nonumber \\
		& = \sum \nolimits_{d=1}^D b_{e,q,n,d}f_{t,n,d},
	\end{align}
	where $b_{e,q,n,d} = \sqrt{\frac{\sigma_d\beta_0G_0}{4\pi}} r_{e,d,q}^{-1} r_{n,d}^{-1} e^{-j\frac{2\pi}{\lambda}(r_{n,d}+r_{e,d,q})+j\chi_d}$ and $r_{e,d,q} = \Vert \mathbf{s}_d - \mathbf{e}_{q}\Vert_2$ is the cluster-to-eavesdropper antenna $q$ distance. 
	Then, we have
	\begin{align}
		h_{e,q,n} = h_{e,q,n}^{\mathrm{LoS}}(\mathbf{f}_{t,n})+h_{e,q,n}^{\mathrm{NLoS}}(\mathbf{f}_{t,n}). 
	\end{align}
	The overall multipath channel between the transmitter and the  eavesdropper is
	\begin{align}
		\label{channel_h}
		&\!\!\! \mathbf{H}_e(\mathbf{f}_{t}) \!=\! [\hat{\mathbf{h}}_{e,1}(\mathbf{f}_{t,1}), \hat{\mathbf{h}}_{e,2}(\mathbf{f}_{t,2}), \ldots, \hat{\mathbf{h}}_{e,N}(\mathbf{f}_{t,N})] \!\in \!\mathbb C^{Q\times N},
	\end{align}
	where $\hat{\mathbf{h}}_{e,n}(\mathbf{f}_{t,n}) \!=\! [h_{e,1,n}, h_{e,2,n}, \ldots, h_{e,Q,n}]^\mathrm{T} \in \mathbb C^{Q\times 1}$.
	
	Let $\mathbf{s} \in \mathbb{C}^{d}$ denote the data symbol vector with $\mathbb{E}\{\mathbf{s}\mathbf{s}^H\} =\mathbf{I}_d$, where $d$ is the number of data streams. Define $\mathbf{W} \in \mathbb{C}^{N \times d} $ as the precoder.  The received signal vector at the receiver is given by
	\begin{align}
		\mathbf{y} = \mathbf{H}(\mathbf{W} \mathbf{s} + \mathbf{z})+\mathbf{n}_0,
	\end{align}
	where $\mathbf{z}$ is the AN used to interfere with the eavesdropper, satisfying $\mathbf{z} \sim \mathcal{CN}(0,\mathbf{R}_z)$. Let $\mathbf{n}_0 \sim \mathcal{CN}(0, \sigma_0^2 \mathbf{I}_M)$ denote the additive white Gaussian noise (AWGN) vector at the receiver.
	Then, the achievable rate of the receiver is
	\begin{align}
		& \!\!\!\!\!\!\!\! R \!= \! \log_2\! \det  \left( \mathbf{I}_{M} \!+\! \mathbf{H} \mathbf{W} \mathbf{W}^H \mathbf{H}^H  \bm{\Sigma}_0^{-1} \right),
	\end{align}
	where $\bm{\Sigma}_0 = \mathbf{H} \mathbf{R}_z \mathbf{H}^H + \sigma_0^2 \mathbf{I}_{M}$.
	Then, the received signal at the eavesdropper is given by
	\begin{align}
		\mathbf{y}_e = \mathbf{H}_e(\mathbf{W} \mathbf{s} + \mathbf{z})+\mathbf{n}_e.
	\end{align}
	Thus, the rate of the wiretap channel for the message $\mathbf{s}$ at the eavesdropper is given by
	\begin{align}
		&\!\!\!\!\!\!\!\!R_{e} \!= \! \log_2 \! \det  \left( \mathbf{I}_{Q} \!+\! \mathbf{H}_e \mathbf{W} \mathbf{W}^H \mathbf{H}_e^H \bm{\Sigma}_e^{-1} \right),
	\end{align}
	where $\bm{\Sigma}_e = \mathbf{H}_e \mathbf{R}_z \mathbf{H}_e^H +  \sigma_e^2 \mathbf{I}_{Q}$. 
	The corresponding SR is given by $R_s = [{R} - R_{e}]_{+}$.
	
	\vspace{-4mm}
	\subsection{Problem Formulation}
	\vspace{-1mm}
	In this paper, we formulate a SR maximization problem by jointly optimizing the transmit precoding matrix $\mathbf{W}$, the AN covariance matrix $\mathbf{R}_z$, and the RA orientations $\mathbf{f}_{t}$, $\mathbf{f}_{r}$, subject to the maximum zenith angle constraints. Firstly, let $\mathbf{R}_z = \mathbf{W}_e\mathbf{W}_e^H$, where $\mathbf{W}_e \in \mathbb{C}^{N \times N}$. Accordingly, the related optimization problem is given by
	\begin{subequations}
		\label{P0}
		\begin{eqnarray}
			\label{P0-0}
			&\!\!\!\!\!\!\!\!\!\!\!\!\!\!\!\!\! \max  \limits_{\mathbf{f}_{t},\mathbf{f}_{r}, \mathbf{W},\mathbf{W}_e}  
			&\!\!\!\!\!\! R_{\mathrm{s}} \\
			\label{P0-1}
			&\!\!\!\!\!\!\!\!\!\!\!\!\!\!\!\!\! \mathrm{s.t.}  &\!\!\!\!\!\! \mathrm{Tr}(\mathbf{W}\mathbf{W}^H) + \mathrm{Tr}(\mathbf{W}_e\mathbf{W}_e^H) \leq P_{\max}, \\
			\label{P0-2}
			&&\!\!\!\!\!\! \cos(\theta_{\max}) \le \mathbf{f}_{t,n}^T\mathbf{e}_z \le 1,  \forall n \in \mathcal{N},\\
			\label{P0-3}
			&&\!\!\!\!\!\! \cos(\theta_{\max}) \le \mathbf{f}_{r,m}^T\mathbf{e}_z \le 1, \forall m \in \mathcal{M},\\
			\label{P0-4}
			&&\!\!\!\!\!\! \Vert \mathbf{f}_{t,n} \Vert_2 = 1, \Vert \mathbf{f}_{r,m} \Vert_2 = 1, \forall n \in \mathcal{N}, \forall m \in \mathcal{M},
		\end{eqnarray}
	\end{subequations}
	where $P_{\max}$ denotes the maximum power budget at the transmitter.
	Problem $(\mathrm{\ref{P0}})$ is challenging to solve due to the highly non-concave objective function with respect to the RA orientations and the non-convex maximum zenith angle constraints in $(\mathrm{\ref{P0-2}})$ and $(\mathrm{\ref{P0-3}})$. Therefore, problem $(\mathrm{\ref{P0}})$ is a highly coupled and non-convex optimization problem with manifold-type geometric constraints, for which obtaining a globally optimal solution is generally intractable. To tackle it, we develop an efficient AO algorithm that decomposes the original problem into a sequence of more tractable subproblems and alternately optimizes the transmit beamforming, AN design, and RA orientations.
	
	\vspace{-4mm}
	\section{SISO case}   \label{Single antenna case}
	\vspace{-2mm}
	In this section, we study a simplified SISO LoS-dominant case to provide analytical 
	insights into how RA orientation affects the tradeoff between legitimate-link 
	enhancement and eavesdropping-link suppression. 
	Specifically, we set $M=N=Q=1$ without AN. To focus on the directional optimization, we consider the LoS-dominant case by setting $D=0$. Under this simplification, the legitimate and eavesdropping channels reduce to $h = af_{t,r} f_{r,t}$ and $h_e = a_ef_{t,e}$,
	where $a$ and $a_e$ collect the distance-dependent propagation factors and phases. We have
	\begin{equation}
		f_{t,r} = [\tilde{\mathbf f}_t^T \mathbf u_{tr}]_+^p,
		f_{r,t} = [\tilde{\mathbf f}_r^T \mathbf u_{rt}]_+^p,
		f_{t,e} = [\tilde{\mathbf f}_t^T \mathbf u_{te}]_+^p,
	\end{equation}
	where $\tilde{\mathbf f}_t = \mathbf{R}_t\mathbf{f}_{t}$ and $\tilde{\mathbf f}_r = \mathbf{R}_r\mathbf{f}_{r}$ denote the global transmit and receive boresight vectors, respectively, and
	\begin{align}
		& \!\!\! \mathbf u_{tr} = \frac{\mathbf r-\mathbf t}{\|\mathbf r-\mathbf t\|},
		\mathbf u_{rt} = -\mathbf u_{tr},
		\mathbf u_{te} = \frac{\mathbf e-\mathbf t}{\|\mathbf e-\mathbf t\|}.
	\end{align}
	Then, the SR can be written as
	\begin{equation}
		R_s =\left[
		\log_2\!\left(1+\frac{|h w|^2}{\sigma_0^2}\right)
		-\log_2\!\left(1+\frac{|h_e w|^2}{\sigma_e^2}\right)
		\right]_+ .
	\end{equation}
	For fixed channel coefficients $h$ and $h_e$, the optimal transmit coefficient is
	\begin{equation}
		w^\star =
		\begin{cases}
			\sqrt{P_{\max}} e^{j\phi}, & {|h|^2}/{\sigma_0^2} > {|h_e|^2}/{\sigma_e^2}, \\[1ex]
			0, & {|h|^2}/{\sigma_0^2} \le {|h_e|^2}/{\sigma_e^2},
		\end{cases}
	\end{equation}
	where $\phi$ is an arbitrary phase. Hence, whenever transmission is beneficial, the orientation design reduces to
	\begin{equation}
		\max_{\mathbf f_t,\mathbf f_r} \ \ 
		\log_2\!\left(1+\gamma_b f_{t,r}^2 f_{r,t}^2\right)
		-\log_2\!\left(1+\gamma_e f_{t,e}^2\right),
		\label{eq:orientation_problem}
	\end{equation}
	where $\gamma_b \triangleq {P_{\max}|a|^2}/{\sigma_0^2}$ and $
	\gamma_e \triangleq {P_{\max}|a_e|^2}/{\sigma_e^2}$.
	
	We first optimize the receive direction. Since $\mathbf f_r$ only affects the legitimate-link gain $f_{r,t}$, the
	receive RA should align with the incoming LoS direction $\mathbf u_{rt}$.
	Without the zenith angle constraint, this gives 
	\begin{equation}
		\tilde{\mathbf f}^{\star}_r = \mathbf u_{rt},\mathbf f^\star_r=\mathbf R_r^T\mathbf u_{rt}, \rightarrow f^\star_{r,t}=1.
	\end{equation}
	If the zenith angle constraint is imposed, the feasible spherical cap is
	$\mathcal C_r
	\triangleq
	\left\{
	\mathbf f \in \mathbb R^3 :
	\|\mathbf f\|=1,
	\mathbf f^T \mathbf e_z \ge \cos(\theta_{\max})
	\right\}$.
	Then, the constrained-optimal receive direction is the Euclidean projection of $\mathbf R_r^T \mathbf u_{rt}$ onto $\mathcal C_r$, given by $\mathbf f_r^\star = \Pi_{\mathcal C_r}(\mathbf R_r^T \mathbf u_{rt})$,
	with
	\begin{equation}
		\!\!\!\! \Pi_{\mathcal C_r}(\mathbf x) \!=\!
		\begin{cases}
			\mathbf x, & \mathbf x^T\mathbf e_z \!\ge \! c_z,\\[2mm]
			c_z\mathbf e_z \!+ \! s_z\dfrac{\mathbf x_\perp}{\|\mathbf x_\perp\|_2},
			& \mathbf x^T\mathbf e_z \!<\! c_z,\ \|\mathbf x_\perp\|_2 \!> \!0,\\
			c_z\mathbf e_z \!+ \! s_z\mathbf u,
			& \mathbf x^T\mathbf e_z \!<\! c_z,\ \|\mathbf x_\perp\|_2\!=\! 0,
		\end{cases}
	\end{equation}
	where $\mathbf x=\mathbf R_r^T\mathbf u_{rt}$, $c_z=\cos(\theta_{\max})$, $s_z=\sin(\theta_{\max})$, and $\mathbf x_\perp=\mathbf x-(\mathbf x^T\mathbf e_z)\mathbf e_z$. The term $\mathbf{u}$ is any unit vector orthogonal to $\mathbf e_z$.
	Substituting the optimal receive gain $f_{r,t}^\star$ into \eqref{eq:orientation_problem}, we obtain the reduced transmit-direction problem
	\begin{equation}
		\max_{\mathbf f_t}
		J(\tilde{\mathbf f}_t)
		\triangleq
		\log_2\!\left(1\!+\!\bar{\gamma}_b [\tilde{\mathbf f}_t^T \mathbf u_{tr}]_+^{2p}\right)
		\!-\!\log_2\!\left(1\!+\!\gamma_e [\tilde{\mathbf f}_t^T \mathbf u_{te}]_+^{2p}\right),
		\label{eq:reduced_tx_problem}
	\end{equation}
	where $\bar{\gamma}_b \triangleq \gamma_b (f_{r,t}^\star)^2$.
	
	Let $\mathcal S \triangleq \operatorname{span}\{\mathbf u_{tr}, \mathbf u_{te}\}$.
	The objective function in \eqref{eq:reduced_tx_problem} depends on $\tilde{\mathbf f}_t$ through  $\tilde{\mathbf f}_t^T \mathbf u_{tr}$ and $\tilde{\mathbf f}_t^T \mathbf u_{te}$.
	Therefore, any component of $\tilde{\mathbf f}_t$ orthogonal to $\mathcal S$ consumes norm but does not improve either the useful signal gain or the suppression of the eavesdropper gain. Consequently, without the zenith angle constraint, there exists an optimal transmit direction that lies in the plane $\mathcal S$, i.e., $
	\exists \tilde{\mathbf f}_t^\star \in \mathcal S$.
	When the local zenith angle constraint is imposed, 
	since $
	(\mathbf R_t^T\tilde{\mathbf f}_t)^T\mathbf e_z
	=
	\tilde{\mathbf f}_t^T\mathbf R_t\mathbf e_z
	=
	\tilde{\mathbf f}_t^T\mathbf a_t
	\ge \cos(\theta_{\max})$,
	the feasible set 
	also depends on the transformed cap axis \(\mathbf a_t=\mathbf R_t\mathbf e_z\). 
	Therefore, the above planar reduction is globally exact when 
	\(\mathbf a_t\in\mathcal S\), or when the zenith angle constraint is inactive 
	at the optimum. For a general array posture with 
	\(\mathbf a_t\notin\mathcal S\), the following 1D planar search 
	serves as a useful structural characterization of the tradeoff between 
	enhancing the legitimate link and suppressing the eavesdropping link.
	
	Denote $\psi \triangleq \arccos(\mathbf u_{tr}^T \mathbf u_{te})$. When $\mathbf u_{tr}$ and $\mathbf u_{te}$ are collinear, the problem degenerates to a 1D alignment case and can be handled separately. Specifically, if $\psi=0$, the legitimate receiver and
	the eavesdropper are located in the same transmit direction. In this case,
	orientation rotation cannot separate the two links in the angular domain.
	The optimal strategy is to maximize the common directional gain when
	$\bar{\gamma}_b>\gamma_e$, and no positive secrecy rate can be obtained
	otherwise. If $\psi=\pi$, steering the transmit boresight toward the
	legitimate receiver naturally places the eavesdropper in the backward
	half-space, so the eavesdropper gain is nulled. Then, we focus on the non-collinear case $\psi  \in (0,\pi)$.
	We first construct an orthonormal basis of $\mathcal S$ as
	\begin{equation}
		\mathbf v_1 = \mathbf u_{tr},
		\mathbf v_2 =
		\frac{\mathbf u_{te} - (\mathbf u_{tr}^T \mathbf u_{te})\mathbf u_{tr}}
		{\left\|\mathbf u_{te} - (\mathbf u_{tr}^T \mathbf u_{te})\mathbf u_{tr}\right\|}.
	\end{equation}
	The transmit boresight within $\mathcal S$ can be expressed by $\theta$
	\begin{equation}
		\tilde{\mathbf f}_t(\theta)=\cos\theta \mathbf v_1-\sin\theta\mathbf v_2.
		\label{eq:tx_param}
	\end{equation}
	Then, we have $\tilde{\mathbf f}_t(\theta)^T \mathbf u_{tr} = \cos\theta$.
	Since $\mathbf u_{te} = \cos\psi \mathbf v_1 + \sin\psi \mathbf v_2$,
	we get
	\begin{equation}
		\tilde{\mathbf f}_t(\theta)^T \mathbf u_{te}
		=\cos\psi \cos\theta - \sin\psi \sin\theta
		=\cos(\psi+\theta).
	\end{equation}
	Since $\tilde{\mathbf f}_t(\theta)^T\mathbf u_{tr}=\cos\theta$, the useful-link gain is nonzero only when $\cos\theta \ge 0$. Moreover, under the chosen sign convention of $\mathbf v_2$, a positive $\theta$ corresponds to rotating the transmit boresight away from the eavesdropper side.  
	To characterize the tradeoff of steering away from the eavesdropper while 
	maintaining a nonzero useful-link gain, we focus on the interval $\theta \in [0, \pi/2]$. 
	Thus, the corresponding transmit directional gains are
	\begin{equation}
		f_{t,r}(\theta)=\cos^p\theta,
		f_{t,e}(\theta)=[\cos(\psi+\theta)]_+^p.
	\end{equation}
	Define the nulling angle $\theta_0 \triangleq \max\{0,\frac{\pi}{2}-\psi\}$.
	For $\theta\ge \theta_0$, we have $\cos(\psi+\theta)\le 0$, so the eavesdropper gain is completely nulled. Meanwhile, the
	legitimate receiver still obtains a nonzero transmit gain $f_{t,r}(\theta_0)=\cos^p\theta_0$.
	For $\psi\in(0,\pi/2]$, this becomes $f_{t,r}(\theta_0)=\sin^p\psi$. Define $\Theta_{\mathrm{search}}= \Theta_{\mathrm{feas}} \cap [0, \pi/2] $ as the feasible interval induced by the zenith angle constraint on $\mathbf f_t(\theta)=\mathbf R_t^T \tilde{\mathbf f}_t(\theta)$, where
	\begin{equation}
		\label{theta_feasible}
		\Theta_{\mathrm{feas}}
		=\left\{\theta:(\mathbf R_t^T \tilde{\mathbf f}_t(\theta))^T \mathbf e_z\ge\cos(\theta_{\max})
		\right\}.
	\end{equation}
	Substituting \eqref{eq:tx_param}, we obtain
	\begin{equation}
		(\mathbf R_t^T \tilde{\mathbf f}_t(\theta))^T \mathbf e_z=a\cos\theta - b\sin\theta,
	\end{equation}
	where $a = \mathbf v_1^T \mathbf R_t \mathbf e_z,
	b = \mathbf v_2^T \mathbf R_t \mathbf e_z$.
	Thus, the feasible set is reformulated as $
	\Theta_{\mathrm{feas}}=\left\{
	\theta:a\cos\theta - b\sin\theta
	\ge\cos(\theta_{\max})\right\}$.
	The resulting SR achieved by the leakage-nulling
	design is
	\begin{equation}
		R_s^{\mathrm{LN}}
		=
		\log_2\left(1+\bar{\gamma}_b\cos^{2p}\theta_0\right),
	\end{equation}
	which is strictly positive if $\bar{\gamma}_b>0$ and the feasible leakage-nulling angle satisfies $\theta_0\in\Theta_{\rm search}$ with $\theta_0<\pi/2$.
This result reveals an important difference between RA-aided and conventional
fixed-orientation SISO wiretap channels. While an unfavorable channel ordering
	${|h|^2}/{\sigma_0^2}\leq {|h_e|^2}/{\sigma_e^2}$ leads to zero secrecy
	capacity in the conventional case, RA rotation can reshape the effective
	gains by forcing $f_{t,e}=0$ while preserving $f_{t,r}>0$. Hence, it can
	convert an unfavorable wiretap channel into a positive-secrecy channel.
	Nevertheless, this
	leakage-nulling rule is not necessarily globally optimal, since enforcing
	$f_{t,e}=0$ may cause a large legitimate-link gain loss when $\psi$ is small.
	This motivates the following 1D search for a better transmit
	orientation within the planar characterization. 	To obtain the optimal planar transmit orientation, we further reformulate problem \eqref{eq:reduced_tx_problem} into
		\begin{align}
		 \max_{\theta \in \Theta_{\mathrm{search}}}
		\!\! J(\theta)\!=\!\log_2\!\left(1 \!\!+\!\! \bar{\gamma}_b \cos^{2p}\theta\right) 
		\!-\! \log_2 \left(1 \!\!+\!\! \gamma_e \cos^{2p}(\psi \!+ \! \theta)\right).\nonumber
	\end{align}
		Then, we perform a 1D search over $\Theta_{\rm search}$ to obtain $\theta^\star=\arg\max_{\theta\in\Theta_{\rm search}}J(\theta)$.
	Once $\theta^\star$ is obtained, the optimal global transmit boresight is $\tilde{\mathbf f}_t^\star=\cos\theta^\star\,\mathbf v_1
	-\sin\theta^\star\,\mathbf v_2$,
	and the corresponding local transmit direction is $\mathbf f_t^\star = \mathbf R_t^T \tilde{\mathbf f}_t^\star$. 

\begin{figure}[t]
		\centering
		\includegraphics[width=0.49 \textwidth]{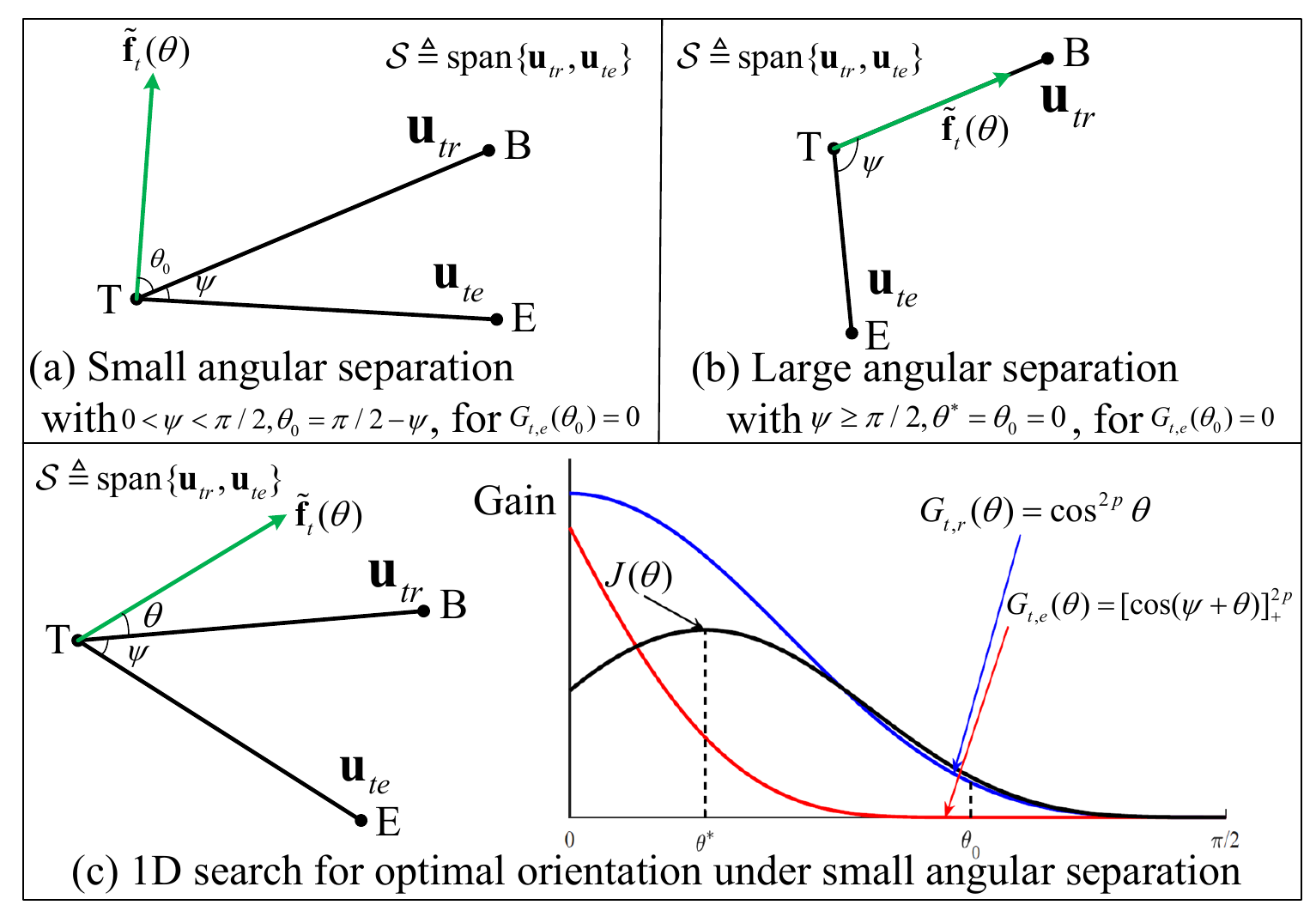}
		\caption{SISO secrecy-oriented RA rotation under the unconstrained planar characterization in
			$\mathcal S$. The nulling angle
			$\theta_0=\max\{0,\pi/2-\psi\}$ guarantees zero eavesdropper gain if feasible, while the
			1D search finds the balanced rotation angle $\theta^\star$
			that maximizes the SR.}
		\label{fig:siso_insight}
		\vspace{-6mm}
	\end{figure}

	Fig.~\ref{fig:siso_insight}
	shows the geometric interpretation of the secrecy-oriented RA rotation,
	where $T$, $E$, and $B$ denote the transmitter, eavesdropper, and legitimate
	receiver, respectively. For clarity, Fig.~\ref{fig:siso_insight} is drawn
	for the unconstrained planar characterization in
	$\mathcal S\triangleq \mathrm{span}\{\mathbf u_{tr},\mathbf u_{te}\}$.
	In this plane, rotating the transmit boresight away from eavesdropper yields the
	directional gains
	$G_{t,r}(\theta)=\cos^{2p}\theta$ and
	$G_{t,e}(\theta)=[\cos(\psi+\theta)]_+^{2p}$. Fig.~\ref{fig:siso_insight}(a) shows the case $0<\psi<\pi/2$, where the nulling angle
	$\theta_0=\pi/2-\psi$ places eavesdropper outside the transmit front half-space. 
	Fig.~\ref{fig:siso_insight}(b) presents the case $\psi\geq\pi/2$, where directly aligning the
	boresight with legitimate direction already nulls eavesdropper, yielding $\theta^\star=0$.
	Fig.~\ref{fig:siso_insight}(c) illustrates the 1D search of $J(\theta)$ to obtain the optimal $\theta^{\ast}$ for the case $0<\psi<\pi/2$, where $\theta^\star$ balances legitimate-link gain and leakage suppression.
	Practical zenith angle constraints are incorporated through
	$\Theta_{\rm search}$ in \eqref{theta_feasible}.
	
	The above SISO analysis reveals that RA rotation introduces an angular-domain
	secrecy DoF by reshaping the effective legitimate and eavesdropping gains.
	While this mechanism can be characterized by a single rotation angle in the
	SISO case, it becomes a coupled matrix-level channel reshaping problem in the
	general MIMO case. Specifically, each transmit RA simultaneously affects
	multiple entries of both the legitimate and eavesdropping channel matrices,
	and the RA orientations are further coupled with transmit beamforming and AN
	design. This motivates the AO-based joint optimization framework developed in
	Section~\ref{Proposed Solution}.

	\vspace{-4mm}
	\section{MIMO Case}  \label{Proposed Solution}
	\vspace{-2mm}
	In this section, we first reformulate the objective function of problem \eqref{P0} into a more tractable form. Then, an AO algorithm is applied to alternately optimize the transmit matrix $\mathbf{W}$, the decomposition of AN covariance matrix $\mathbf{W}_e$, and the RA orientations $\mathbf{f}_{t}$, $\mathbf{f}_{r}$.
	
	\vspace{-4mm}
	\subsection{Problem Reformulation}
	\vspace{-2mm}
	The SR in \eqref{P0-0} can be expressed as
	\begin{align}
		R_{\mathrm{s}} =&\log_2 \det  \left( \mathbf{I}_{M} + \mathbf{H} \mathbf{W} \mathbf{W}^H \mathbf{H}^H  \bm{\Sigma}_0^{-1} \right)  \nonumber \\
		&  -\log_2 \! \det  \left( \mathbf{I}_{Q} \!+\! \mathbf{H}_e \mathbf{W} \mathbf{W}^H \mathbf{H}_e^H \bm{\Sigma}_e^{-1} \right)   \nonumber \\
		=& \underbrace{\log_2 \det \left( \mathbf{I}_{M} + \mathbf{H} \mathbf{W} \mathbf{W}^H \mathbf{H}^H \bm{\Sigma}_0^{-1} \right)}_{f_1} \nonumber \\
		& + \underbrace{\log_2 \det \left( \mathbf{I}_{Q} + \mathbf{H}_e \mathbf{W}_e \mathbf{W}_e^H \mathbf{H}_e^H (\sigma_e^2 \mathbf{I}_{Q})^{-1} \right)}_{f_2} \nonumber \\
		& - \underbrace{\log_2 \det \left( \mathbf{I}_{Q} + \sigma_e^{-2}\mathbf{H}_e (\mathbf{W} \mathbf{W}^H + \mathbf{W}_e \mathbf{W}_e^H) \mathbf{H}_e^H \right)}_{f_3}, \nonumber
	\end{align}
	which remains non-convex and difficult to solve. Therefore, we reformulate the expressions of $f_1$, $f_2$ and $f_3$ into more tractable forms. In terms of $f_1$ and $f_2$, we employ the MMSE-based transformation, which converts the original rate expressions into equivalent and more tractable formulations by introducing auxiliary matrices. Specifically, for $f_{1}$, we introduce $\mathbf{U} \in \mathbb{C}^{M \times d}$ as the linear decoding matrix to estimate the data streams, i.e., $\hat{\mathbf{s}} = \mathbf{U}^H \mathbf{y}$. Then, the mean-square error (MSE) matrix of estimation is given by
	\begin{align}
		\mathbf{E} &= \E\left[ (\hat{\mathbf{s}} - \mathbf{s})(\hat{\mathbf{s}} - \mathbf{s})^H \right] \nonumber \\
		&= \mathbf{I} - \mathbf{U}^H \mathbf{H} \mathbf{W} - \mathbf{W}^H \mathbf{H}^H \mathbf{U} \nonumber \\
		&
		+ \mathbf{U}^H \left(\mathbf{H} \mathbf{W} \mathbf{W}^H \mathbf{H}^H +\mathbf{H} \mathbf{W}_e \mathbf{W}_e^H \mathbf{H}^H+ \sigma_0^2 \mathbf{I} \right) \mathbf{U}. \label{eq:complete_Ek} 
	\end{align}
	By introducing an auxiliary matrix $\mathbf{\Omega} \in \mathbb{C}^{d \times d}$, $f_{1}$ can be reformulated as \cite{Shi2015}
	\begin{align}
		&f_{1} = {1}/{\mathrm{ln2}}\max_{\mathbf{U}, \mathbf{\Omega}} h_{1} (\mathbf{U}, \mathbf{\Omega}, \mathbf{W},\mathbf{W}_e), \\
		&h_{1} (\mathbf{U}, \mathbf{\Omega}, \mathbf{W},\mathbf{W}_e) \triangleq  \ln \det |\mathbf{\Omega}| - \text{Tr} (\mathbf{\Omega} \mathbf{E}) + d. 
	\end{align}
	Note that $h_{1} (\mathbf{U}, \mathbf{\Omega}, \mathbf{W})$ is concave w.r.t. $\mathbf{U}, \mathbf{\Omega}$ with other variables fixed. By considering the first-order optimality condition, the optimal solutions of $\mathbf{U}$ and $ \mathbf{\Omega}$ can be derived as
	\begin{align}
		\label{U_update}
		&\!\mathbf{U}^{\ast} \!=\! \left(  \mathbf{H} \mathbf{W} \mathbf{W}^H \mathbf{H}^H \!+\!\mathbf{H} \mathbf{W}_e \mathbf{W}_e^H \mathbf{H}^H \!+\! \sigma_0^2 \mathbf{I} \right)^{-1} \mathbf{H} \mathbf{W}, \\
		\label{Omega_update}
		&\!\mathbf{\Omega}^{\ast} \!=\! \mathbf{E}^{-1}.
	\end{align}
	Similarly, for $f_{2}$, we introduce an auxiliary linear receiver $\mathbf{U}_{e} \in \mathbb{C}^{Q \times N}$ associated with the equivalent AN covariance term. The MSE matrix for estimation is given by
	\begin{align}
		\mathbf{E}_{e} =& \mathbf{I} - \mathbf{U}_{e}^H \mathbf{H}_e \mathbf{W}_e - \mathbf{W}_e^H \mathbf{H}_e^H \mathbf{U}_{e} +\nonumber \\
		& \mathbf{U}_{e}^H \left( \mathbf{H}_e \mathbf{W}_e \mathbf{W}_e^H \mathbf{H}_e^H + \sigma_e^2 \mathbf{I} \right) \mathbf{U}_{e}. \label{eq:complete_Ejk}
	\end{align}
	Then, by introducing $\mathbf{\Omega}_e \in \mathbb{C}^{N \times N}$, $f_{2}$ can be reformulated as
	\begin{align}
		&f_{2} = {1}/{\mathrm{ln2}}\max_{\mathbf{U}_{e}, \mathbf{\Omega}_{e}} h_{2} (\mathbf{U}_{e}, \mathbf{\Omega}_{e}, \mathbf{W}_e),\\
		\label{h2}
		&h_{2} (\mathbf{U}_{e}, \mathbf{\Omega}_{e}, \mathbf{W}_e) \triangleq \ln \det |\mathbf{\Omega}_{e}| - \text{Tr} (\mathbf{\Omega}_{e} \mathbf{E}_{e}) + N. 
	\end{align}
	The optimal solutions of $\mathbf{U}_{e}$ and $ \mathbf{\Omega}_{e}$ can be derived as
	\begin{align}
		\label{Ue_update}
		& \mathbf{U}_{e}^{\ast} = \left( \mathbf{H}_e \mathbf{W}_e \mathbf{W}_e^H \mathbf{H}_e^H + \sigma_e^2 \mathbf{I} \right)^{-1} \mathbf{H}_e \mathbf{W}_e, \\
		\label{Omegae_update}
		&\mathbf{\Omega}_{e}^{\ast} = \mathbf{E}_{e}^{-1}.
	\end{align}
	Then, we present a useful lemma \cite{ChristensenLemma2008}, which enables the introduction of auxiliary variables to obtain solvable subproblems.
	\begin{lemma}
		\label{lemma1}
		\rm{Let $\mathbf{E} \in \mathbb{C}^{N \times N}$ be any matrix such that $\mathbf{E} \succeq \mathbf{0}$. Since $f(\mathbf{W}) = -\mathrm{Tr}(\mathbf{WE}) + \ln |\mathbf{W}| + N$, we have
			\begin{equation}
				\ln |\mathbf{E}^{-1}| = \max_{\mathbf{W} \succeq \mathbf{0}} f(\mathbf{W}),
			\end{equation}
			and the optimal solution is given by $\mathbf{W}^{\star} = \mathbf{E}^{-1}$.}
	\end{lemma}
	\vspace{-3mm}
	According to Lemma \ref{lemma1}, define 
	\begin{align}
		& \mathbf{E}_x =  \left( \mathbf{I}_{Q} \!+\!{\sigma}_e^{-2}\mathbf{H}_e (\mathbf{W} \mathbf{W}^H+\mathbf{W}_e \mathbf{W}_e^H) \mathbf{H}_e^H  \right).
	\end{align}
	Then, we introduce an auxiliary matrix $\mathbf{\Omega}_x \in \mathbb{C}^{Q \times Q}$. $f_{3}$ can be reformulated as 
	\begin{align}
		&-f_{3} = {1}/{\mathrm{ln2}}\max_{\mathbf{\Omega}_{x}} h_{3} ( \mathbf{\Omega}_{x}, \mathbf{W},\mathbf{W}_e),\\
		\label{h3}
		&h_{3} ( \mathbf{\Omega}_{x}, \mathbf{W},\mathbf{W}_e) \triangleq \ln \det |\mathbf{\Omega}_{x}| - \text{Tr} (\mathbf{\Omega}_{x} \mathbf{E}_{x}) + Q, 
	\end{align}
	where the optimal solution is given  by 
	\begin{align}
		\label{Omegax_update}
		\mathbf{\Omega}_x^{\ast} = \mathbf{E}_{x}^{-1}.
	\end{align}
	Based on the above, the objective function \eqref{P0-0} can be rewritten as  $
	R_{\mathrm{s}} = {1}/{\mathrm{ln2}}\max_{\mathbf{K}} F(\mathbf{K}, \mathbf W,\mathbf W_e,\mathbf f_t,\mathbf f_r)$, 
	where $\mathbf{K} \triangleq [\mathbf{U},\mathbf{\Omega},\mathbf{U}_{e}, \mathbf{\Omega}_{e},\mathbf{\Omega}_{x}] $ and $F(\mathbf{K}, \mathbf W,\mathbf W_e,\mathbf f_t,\mathbf f_r) = h_{1} (\mathbf{U}, \mathbf{\Omega}, \mathbf{W},\mathbf{W}_e) + h_{2} (\mathbf{U}_{e}, \mathbf{\Omega}_{e}, \mathbf{W}_e) + h_{3} ( \mathbf{\Omega}_{x}, \mathbf{W},\mathbf{W}_e)$.
	Since ${1}/{\mathrm{ln2}}$ is a common positive constant, we neglect it for simplicity. Then, problem $(\mathrm{\ref{P0}})$ is rewritten as 
	\begin{subequations}
		\label{P1}
		\begin{eqnarray}
			\label{P1-0}
			&\!\!\!\!\!\!\!\!\!\!\!\!\!\!\!\!\! \max  \limits_{\mathbf{f}_{t},\mathbf{f}_{r}, \mathbf{W}_e, \mathbf{W},\mathbf{K}}  
			&\!\!\!\!\!\! F(\mathbf{K},\mathbf{W}_e, \mathbf{W},\mathbf{f}_{t},\mathbf{f}_{r}) \\
			\label{P1-1}
			&\!\!\!\!\!\!\!\!\!\! \mathrm{s.t.}  &\!\!\!\!\!\! \eqref{P0-1},\eqref{P0-2},\eqref{P0-3},\eqref{P0-4},\\
			\label{P1-4}
			&&\!\!\!\!\!\! \mathbf{\Omega} \succeq \mathbf{0},\mathbf{\Omega}_e \succeq \mathbf{0},\mathbf{\Omega}_x \succeq \mathbf{0}.
		\end{eqnarray}
	\end{subequations}
	Note that the objective function \eqref{P1-0} is concave w.r.t. $\{\mathbf{U},\mathbf{\Omega},\mathbf{U}_{e}, \mathbf{\Omega}_{e},\mathbf{\Omega}_{x}\}$ with fixed $\{\mathbf{W},\mathbf{W}_e,\mathbf{f}_{t},\mathbf{f}_{r}\}$ and concave w.r.t. $\{\mathbf{W},\mathbf{W}_e\}$ with fixed $\{\mathbf{U},\mathbf{\Omega},\mathbf{U}_{e}, \mathbf{\Omega}_{e},\mathbf{\Omega}_{x},\mathbf{f}_{t},\mathbf{f}_{r}\}$, respectively. We first update $\{\mathbf{U},\mathbf{\Omega},\mathbf{U}_{e}, \mathbf{\Omega}_{e},\mathbf{\Omega}_{x}\}$ in closed forms with fixed $\{\mathbf{W},\mathbf{W}_e,\mathbf{f}_{t},\mathbf{f}_{r}\}$. Then, with given $\{\mathbf{U},\mathbf{\Omega},\mathbf{U}_{e}, \mathbf{\Omega}_{e},\mathbf{\Omega}_{x}\}$, we first update $\{\mathbf{W},\mathbf{W}_e,\mathbf{f}_{t},\mathbf{f}_{r}\}$ by solving the following subproblem
	\begin{subequations}
		\label{P10}
		\begin{eqnarray}
			\label{P20-0}
			&\!\!\!\!\!\!\!\!\!\!\!\!\!\!\!\!\! \max  \limits_{\mathbf{f}_{t},\mathbf{f}_{r}, \mathbf{W}_e, \mathbf{W}}  
			&\!\!\!\!\!\! F(\mathbf{W}_e, \mathbf{W},\mathbf{f}_{t},\mathbf{f}_{r}) \\
			\label{P20-1}
			&\!\!\!\!\!\!\!\!\!\! \mathrm{s.t.}  &\!\!\!\!\!\! \eqref{P0-1},\eqref{P0-2},\eqref{P0-3},\eqref{P0-4}.
		\end{eqnarray}
	\end{subequations}
	
	\vspace{-4mm}
	\subsection{Transmit Beamforming Design}
	\vspace{-2mm}
	Given $\{\mathbf{f}_{t},\mathbf{f}_{r}\}$, we optimize $\mathbf{W}$ and $\mathbf{W}_e$ in this subsection.
	The objective function \eqref{P20-0} can be formulated as 
	\begin{align}
		L(\mathbf{W},\mathbf{W}_e)  
		\!=&\! \mathrm{Tr}\left(\mathbf{W}^H  \widetilde{\mathbf{H}}  \mathbf{W} \right) \!\!-\!\! 2 \Re\left\{ \mathrm{Tr}\left( \mathbf{\Omega} \mathbf{W}^H {\mathbf{H}}^H \mathbf{U} \right) \right\}  \nonumber \\
		& \!+\!\! \mathrm{Tr}\! \left(\! \mathbf{W}_e^H  \widetilde{\mathbf{H}}_e  \mathbf{W}_e \! \right) \!\!-\!\! 2 \Re\left\{\! \mathrm{Tr} \! \left( \mathbf{\Omega}_e \mathbf{W}_e^H {\mathbf{H}}_e^H \mathbf{U}_e \right) \!\right\}, \nonumber
	\end{align}
	where $  \widetilde{\mathbf{H}} \!=\! \mathbf{H}^H\mathbf{U}\mathbf{\Omega}\mathbf{U}^H\mathbf{H} + \sigma_e^{-2}\mathbf{H}_e^H\mathbf{\Omega}_x\mathbf{H}_e$ and $
	\widetilde{\mathbf{H}}_e \!=\! \mathbf{H}^H\mathbf{U}\mathbf{\Omega}\mathbf{U}^H\mathbf{H} \!+\!\mathbf{H}_e^H\mathbf{U}_e\mathbf{\Omega}_e\mathbf{U}_e^H\mathbf{H}_e \!\!+\!\! \sigma_e^{-2}\mathbf{H}_e^H\mathbf{\Omega}_x\mathbf{H}_e$.
	Then, the subproblem regarding $\mathbf{W}$ and $\mathbf{W}_e$ is given by
	\begin{subequations}
		\label{P2}
		\begin{eqnarray}
			\label{P2-0}
			& \min  \limits_{\mathbf{W},\mathbf{W}_e}  
			& L(\mathbf{W},\mathbf{W}_e) \\
			\label{P2-1}
			& \mathrm{s.t.}  & \eqref{P0-1}.
		\end{eqnarray}
	\end{subequations}
	Since the above problem is a convex quadratically constrained quadratic program (QCQP) problem, we apply Lagrangian multiplier method to solve it, which is given by
	\begin{align}
		\label{W_solution}
		&\mathbf{W}^{\ast} = (\widetilde{\mathbf{H}}+\xi \mathbf{I})^{\dagger}(\mathbf{H}^H\mathbf{U}\mathbf{\Omega}),\\
		\label{We_solution}
		&\mathbf{W}_e^{\ast} = (\widetilde{\mathbf{H}}_e+\xi \mathbf{I})^{\dagger}(\mathbf{H}_e^H\mathbf{U}_e\mathbf{\Omega}_e),
	\end{align}
	where $\xi \ge 0$ is the Lagrangian multiplier associated with the transmit power constraint and it can be derived by the bisection search method to satisfy the complementary slackness condition. The detailed derivation is given in Appendix~\ref{Appendix A}.
	
		\begin{algorithm}[t]
		\caption{Riemannian Frank-Wolfe-based Algorithm for Solving Problem (\ref{P3})}
		\label{alg:pointing-vector-design}
		\begin{algorithmic}[1]
			\State \textbf{Input:} Fixed transmit beamforming matrices $\mathbf{W}$ and $\mathbf{W}_e$, 
			auxiliary matrices $\mathbf{U}$, $\mathbf{\Omega}$, $\mathbf{U}_e$, $\mathbf{\Omega}_e$, $\mathbf{\Omega}_x$, 
			initial feasible RA orientations $\{\mathbf{f}_{t,n}^{(0)}\}_{n=1}^{N}$ and $\{\mathbf{f}_{r,m}^{(0)}\}_{m=1}^{M}$,
			maximum zenith angle $\theta_{\max}$, tolerance $\varepsilon$, Armijo parameter $c$, 
			backtracking factor $\beta$, and maximum iteration number $I_{\max}$;
			\State \textbf{Initialize:} Set iteration index $i=0$;
			\Repeat
			\For{$n=1$ to $N$}
			\State Compute the Euclidean gradient  $\nabla_{\mathbf{f}_{t,n}} L$;
			\State Project the gradient onto the tangent space by \eqref{eq:riemannian_gradient};
			\State Solve the Frank-Wolfe linear subproblem by \eqref{eq:fw_linearization};
			\State Set the search direction $
			\mathbf{d}_{t,n}=\mathbf{y}_{t,n}-\mathbf{f}_{t,n}^{(i)}$;
			\State Update $\mathbf{f}_{t,n}^{(i+1)}$ by \eqref{eq:retraction};
			\EndFor
			\For{$m=1$ to $M$}
			\State Compute the Euclidean gradient $\nabla_{\mathbf{f}_{r,m}} L$;
			\State Project the gradient onto the tangent space by \eqref{eq:riemannian_gradient};
			\State Solve the Frank-Wolfe linear subproblem by \eqref{eq:fw_linearization};  
			\State Set the search direction $
			\mathbf{d}_{r,m}=\mathbf{y}_{r,m}-\mathbf{f}_{r,m}^{(i)}$;
			\State Update $\mathbf{f}_{r,m}^{(i+1)}$ by \eqref{eq:retraction};
			\EndFor
			\State Set $i\leftarrow i+1$;
			\Until{$i\ge I_{\max}$ \textbf{or}
				$\left|L^{(i)}-L^{(i-1)}\right|/\left|L^{(i)}\right|\le \varepsilon$};
			\State \textbf{Output:} Converged $\{\mathbf{f}_{t,n}\}_{n=1}^{N}$ and $\{\mathbf{f}_{r,m}\}_{m=1}^{M}$.
		\end{algorithmic}
	\end{algorithm}
	\vspace{-3mm}
	
	\vspace{-2mm}
	\subsection{RA Orientation Design}
	\vspace{-2mm}
	Given $\{\mathbf{W}, \mathbf{W}_e\}$, we optimize $\mathbf{f}_{r}$ and $\mathbf{f}_{t}$ in this subsection. The subproblem is given by
		\begin{eqnarray}
			\label{P3}
			&\!\!\!\!\!\!\!\!\!\! \min  \limits_{\mathbf{f}_{t},\mathbf{f}_{r}}  
			&\!\!\!\! L(\mathbf{f}_t, \mathbf{f}_r) 
			\ \ \  \mathrm{s.t.}  \ \   \eqref{P0-2},\eqref{P0-3},\eqref{P0-4},
		\end{eqnarray}
	where
	\begin{align}
		L(\mathbf{f}_t, \mathbf{f}_r)=&\mathrm{Tr}\left(\mathbf{H} \mathbf{W}_X \mathbf{H}^H \mathbf{C}\right) +\mathrm{Tr}\left(\mathbf{H}_e \mathbf{W}_X \mathbf{H}_e^H \mathbf{C}_X\right) \nonumber \\
		& +\!\mathrm{Tr}\left(\mathbf{H}_e \mathbf{W}_e\mathbf{W}_e^H \mathbf{H}_e^H \mathbf{C}_e\right) \!-\! 2 \Re\left\{ \mathrm{Tr}\left( \mathbf{D} {\mathbf{H}}^H  \right) \right\} \nonumber \\
		& -\! 2 \Re \left\{\mathrm{Tr}\left( \mathbf{D}_e {\mathbf{H}}_e^H  \right)\right\}.
	\end{align}
	The terms $\mathbf{W}_X = \mathbf{W}\mathbf{W}^H + \mathbf{W}_e\mathbf{W}_e^H $,  $\mathbf{C}=\mathbf{U}\mathbf{\Omega}\mathbf{U}^H$, $\mathbf{C}_X=\sigma_e^{-2}\mathbf{\Omega}_x$, $\mathbf{C}_e=\mathbf{U}_e\mathbf{\Omega}_e\mathbf{U}_e^H$, $\mathbf{D}=\mathbf{U}\mathbf{\Omega} \mathbf{W}^H$, and $\mathbf{D}_e=\mathbf{U}_e\mathbf{\Omega}_e \mathbf{W}_e^H$.
	The dependence of $L$ on $(\mathbf f_t,\mathbf f_r)$ is only through $\mathbf H(\mathbf f_t,\mathbf f_r)$ and
	$\mathbf H_e(\mathbf f_t)$.
	Let $[\mathbf H]_{m,n}=h_{m,n}$ and $[\mathbf H_e]_{q,n}=h_{e,q,n}$.
	Define
	\begin{align}
		&\mathbf G \triangleq \frac{\partial L}{\partial \mathbf H^*} = \mathbf C\mathbf H\mathbf W_X - \mathbf D,\\
		&\mathbf G_e \triangleq \frac{\partial L}{\partial \mathbf H_e^*}=\mathbf C_X \mathbf H_e \mathbf W_X
		+ \mathbf C_e \mathbf H_e (\mathbf W_e\mathbf W_e^H)
		-\mathbf D_e.
	\end{align}
	Then, the Euclidean gradients w.r.t.\ the $n$-th transmit RA orientation $\mathbf f_{t,n}$ and the $m$-th receive RA orientation
	$\mathbf f_{r,m}$ follow from elementwise chain rules
	\begin{align}
		&\nabla_{\mathbf f_{t,n}} L
		\!\!=\!\!\!\sum\limits_{m=1}^M \! 2\Re\!\left\{\!
		\mathbf G_{m,n}^{\ast}\frac{\partial h_{m,n}}{\partial \mathbf f_{t,n}}
		\!	\right\} \!\!+\!\!\sum\limits_{q=1}^Q \! 2\Re\!\left\{\!
		(\mathbf G_e)_{q,n}^{\ast}\frac{\partial h_{e,q,n}}{\partial \mathbf f_{t,n}}
		\!\right\}, \nonumber 
		\\
		&\nabla_{\mathbf f_{r,m}} L
		=	\sum\limits_{n=1}^N 2\Re\!\left\{
		\mathbf G_{m,n}^{\ast}\;\frac{\partial h_{m,n}}{\partial \mathbf f_{r,m}}
		\right\}, \nonumber
	\end{align}
	where $G_{m,n}$ denotes the element in the $m$-th row and $n$-th column of matrix $\mathbf{G}$.
	Since
	$h_{m,n}=h^{\mathrm{LoS}}_{m,n}+h^{\mathrm{NLoS}}_{m,n}$
	and
	$h_{e,q,n}=h^{\mathrm{LoS}}_{e,q,n}+h^{\mathrm{NLoS}}_{e,q,n}$, we have
	\begin{align}
		&\!\!\!\!\frac{\partial h_{m,n}}{\partial \mathbf{f}_{t,n}}
		\!\!=\!\!
		\frac{\partial h^{\mathrm{LoS}}_{m,n}}{\partial \mathbf{f}_{t,n}}
		\!+\!
		\frac{\partial h^{\mathrm{NLoS}}_{m,n}}{\partial \mathbf{f}_{t,n}}, 
		\frac{\partial h_{e,q,n}}{\partial \mathbf{f}_{t,n}}
		\!\!=\!\!
		\frac{\partial h^{\mathrm{LoS}}_{e,q,n}}{\partial \mathbf{f}_{t,n}}
		\!+\!
		\frac{\partial h^{\mathrm{NLoS}}_{e,q,n}}{\partial \mathbf{f}_{t,n}},
	\end{align}
	The channel derivations regarding $\frac{\partial h_{m,n}}{\partial \mathbf f_{t,n}}$, $\frac{\partial h_{e,q,n}}{\partial \mathbf f_{t,n}}$, and $\frac{\partial h_{m,n}}{\partial \mathbf f_{r,m}}$ are given in Appendix \ref{Appendix B}.

	The orientation vectors for both transmit and receive RAs reside on the unit sphere subject to a spherical-cap constraint, forming the feasible set
	\begin{equation}
		\mathcal{C} = \left\{ \mathbf{f} \in \mathbb{R}^3 : \| \mathbf{f} \|_2 = 1,\; \mathbf{f}^T \mathbf{e}_z \geq \cos(\theta_{\max}) \right\},
		\label{eq:feasible_set}
	\end{equation}
	where $\mathbf{e}_z = [0,0,1]^T$ and $\theta_{\max}$ is the maximum zenith angle. This feasible set is a compact subset of the unit sphere with a boundary. 
	Hence, its first-order feasible directions are characterized by the tangent cone
	\begin{equation}
		T_{\mathbf{f}}\mathcal{C} = \left\{ \mathbf{v} \in \mathbb{R}^3 : \mathbf{f}^T \mathbf{v} = 0,\; \mathbf{e}_z^T \mathbf{v} \geq 0 \text{ if } \mathbf{f}^T \mathbf{e}_z = \cos(\theta_{\max}) \right\}. \nonumber 
		\label{eq:tangent_space}
	\end{equation}
	The Riemannian gradient is obtained by projecting the Euclidean gradient onto the tangent space
	\begin{equation}
		\text{grad}_{\mathbf{f}} L = \left( \mathbf{I} - \mathbf{f} \mathbf{f}^T \right) \nabla_{\mathbf{f}} L.
		\label{eq:riemannian_gradient}
	\end{equation}
	This projection ensures that the update direction remains on the manifold. To handle the spherical-cap constraint efficiently, we employ a Frank-Wolfe approach. At each iteration, we linearize the objective around the current point $\mathbf{f}$ and solve
	\begin{equation}
		\mathbf{y} = \arg\min_{\mathbf{x} \in \mathcal{C}} \langle \text{grad}_{\mathbf{f}} L,\; \mathbf{x} \rangle.
		\label{eq:fw_linearization}
	\end{equation}
	This linear subproblem admits a closed-form solution on the spherical cap. 
	If \(\|\text{grad}_{\mathbf{f}} L\|_2=0\), we set \(\mathbf y=\mathbf f\). Otherwise, let $\tilde{\mathbf{g}} = \text{grad}_{\mathbf{f}} L / \|\text{grad}_{\mathbf{f}} L\|_2$ be the normalized Riemannian gradient. The solution is given by
	\begin{equation}
		\!\!\! \mathbf{y} \!= \!
		\begin{cases}
			-\tilde{\mathbf{g}}, & \text{if } z \leq -c_z, \\
			-\sqrt{1 - c_z^2} \, \mathbf{v} + c_z \mathbf{e}_z, & \text{if } z > - c_z \text{ and } \|\mathbf{v}\|_2 \neq 0, \\
			\sqrt{1 - c_z^2} \, \mathbf{u} + c_z \mathbf{e}_z, & \text{if } z > -c_z \text{ and } \|\mathbf{v}\|_2 = 0,
		\end{cases}
		\label{eq:fw_solution}
	\end{equation}
	where $\mathbf{v} = (\tilde{\mathbf{g}} - z \mathbf{e}_z) / \|\tilde{\mathbf{g}} - z \mathbf{e}_z\|_2$, $c_z = \cos(\theta_{\max})$, and $z = \tilde{\mathbf{g}}^T \mathbf{e}_z$. The term $\mathbf{u}$ is any unit vector orthogonal to $\mathbf{e}_z$. The Frank-Wolfe search direction is then defined as $\mathbf{d} = \mathbf{y} - \mathbf{f}$.
	Given a stepsize $\rho \in (0,1]$, we use a retraction operation to update orientation vectors $\mathbf{f}_t$ and $\mathbf{f}_{r,m}$ while staying on the manifold, which is given by
	\begin{equation}
		\mathbf{f}_{\text{new}} = \frac{\mathbf{f} + \rho \mathbf{d}}{\| \mathbf{f} + \rho \mathbf{d} \|_2}.
		\label{eq:retraction}
	\end{equation}
	The stepsize $\rho$ is selected via an Armijo backtracking line search to ensure a sufficient decrease in the objective function. Specifically, we choose the largest $\rho$ satisfying
	\begin{equation}
		L(\mathbf{f}_{\text{new}}) \leq L(\mathbf{f}) + c \rho \langle \text{grad}_{\mathbf{f}} L, \mathbf{d} \rangle,
		\label{eq:armijo}
	\end{equation}
	where $c \in (0,1)$ is a constant. The detailed procedures are provided in Algorithm \ref{alg:pointing-vector-design}.

		\begin{algorithm}[t]
		\caption{AO Algorithm for Solving Problem (\ref{P0})}
		\label{alg:overall-section4}
		\begin{algorithmic}[1]
			\State \textbf{Input:} Initial feasible $\mathbf{W}^{(0)}$, $\mathbf{W}_e^{(0)}$, 
			$\mathbf{f}_{t}^{(0)}$, $\mathbf{f}_{r}^{(0)}$,
			transmit power budget $P_{\max}$, maximum zenith angle $\theta_{\max}$, tolerance $\varepsilon$, and maximum iteration number $N_{\max}$;
			\State \textbf{Initialize:} Set outer iteration index $n=0$;
			\Repeat
			\State Fixed $\mathbf{W}^{(n)}$, $\mathbf{W}_e^{(n)}$, $\mathbf{f}_{t}^{(n)}$, and $\mathbf{f}_{r}^{(n)}$, update $\mathbf{U}^{(n+1)}$ and $\mathbf{U}_e^{(n+1)}$ by (\ref{U_update}) and (\ref{Ue_update}), respectively; 
			\State Fixed $\mathbf{W}^{(n)}$, $\mathbf{W}_e^{(n)}$, $\mathbf{f}_{t}^{(n)}$, $\mathbf{f}_{r}^{(n)}$, $\mathbf{U}^{(n+1)}$, and $\mathbf{U}_e^{(n+1)}$,  update $\mathbf{\Omega}^{(n+1)}$, $\mathbf{\Omega}_e^{(n+1)}$, and $\mathbf{\Omega}_x^{(n+1)}$ by (\ref{Omega_update}), (\ref{Omegae_update}), and (\ref{Omegax_update}), respectively;
			\State Fixed $\mathbf{f}_{t}^{(n)}$, $\mathbf{f}_{r}^{(n)}$, $\mathbf{U}^{(n+1)}$, $\mathbf{U}_e^{(n+1)}$, $\mathbf{\Omega}^{(n+1)}$, $\mathbf{\Omega}_e^{(n+1)}$, and $\mathbf{\Omega}_x^{(n+1)}$, update the transmit beamforming $\mathbf{W}^{(n+1)}$ and $\mathbf{W}_e^{(n+1)}$ by calculating $(\mathrm{\ref{W_solution}})$ and $(\mathrm{\ref{We_solution}})$, respectively;
			\State Fixed  $\mathbf{W}^{(n+1)}$, $\mathbf{W}_e^{(n+1)}$, $\mathbf{U}^{(n+1)}$, $\mathbf{U}_e^{(n+1)}$, $\mathbf{\Omega}^{(n+1)}$, $\mathbf{\Omega}_e^{(n+1)}$, and $\mathbf{\Omega}_x^{(n+1)}$, update the transmit and receive RA orientations $\mathbf{f}_{t}^{(n+1)}$ and $\mathbf{f}_{r}^{(n+1)}$ by Algorithm \ref{alg:pointing-vector-design};
			\State Set $n\leftarrow n+1$.
			\Until{$n\ge N_{\max}$ \textbf{or}
				$\left|F^{(n)}-F^{(n-1)}\right|/\left|F^{(n)}\right|\le \varepsilon$};
			\State \textbf{Output:} Converged $\mathbf{W}$, $\mathbf{W}_e$, $\mathbf{f}_{t}$, and $\mathbf{f}_{r}$.
		\end{algorithmic}
	\end{algorithm}
	\vspace{-3mm}
	
	\vspace{-3mm}
	\subsection{Complexity and Convergence Analysis}
	\vspace{-2mm}
	The complete procedure for solving problem (\ref{P0}) is presented in Algorithm \ref{alg:overall-section4}, where the auxiliary variables, the transmit beamforming, the decomposition of AN covariance matrix, and transmit/receive antenna RA orientations are alternately optimized until convergence is achieved. Note that the objective value of problem (\ref{P0}) is non-decreasing by alternately optimizing $\mathbf{K}$, $\mathbf{W}$, $\mathbf{W}_e$, $\mathbf{f}_t$, and $\mathbf{f}_r$. Moreover, under the transmit power constraint and the maximum zenith angle constraints, the objective function of problem (\ref{P0}) is upper bounded. Thus, the AO algorithm ensures convergence.
	
	In terms of computational complexity, the complexities of calculating $\mathbf{U}$ and $\mathbf{U}_e$ are given by $\mathcal{O}(M^3)$ and $\mathcal{O}(Q^3)$, respectively. Besides, the complexities of calculating $\mathbf{\Omega}$, $\mathbf{\Omega}_e$, and $\mathbf{\Omega}_x$ are given by $\mathcal{O}(d^3)$ and $\mathcal{O}(N^3)$, and $\mathcal{O}(Q^3)$, respectively. The updates of the transmit beamforming matrices $\mathbf{W}$ and $\mathbf{W}_e$ have a complexity of $\mathcal{O}(\max\{2N^3,2N^2Q\})$. For Algorithm \ref{alg:pointing-vector-design}, the complexity of calculating the gradient of transmit RA orientations is $\mathcal{O}(N(M+Q))$. Similarly, updating all receive RA orientations requires $\mathcal{O}(MN)$ operations. The Frank-Wolfe direction computation incurs linear complexity $\mathcal{O}(N+M)$. 
	For the Armijo backtracking line search, the complexity is  $\mathcal{O}(I_AN(M+Q))$. Thus, the total complexity of Algorithm \ref{alg:pointing-vector-design} can be written as $\mathcal{C}_{FW} = \mathcal{O}\!\left(I_{\mathrm{FW}}\left[2MN+NQ + I_{\mathrm{A}}(NQ+MN)\right]\right)$, where $I_{\mathrm{FW}}$ is the number of iterations of Algorithm \ref{alg:pointing-vector-design}, and $I_{\mathrm{A}}$ is the number of iterations of the Armijo backtracking line search. Therefore, 
	the overall complexity of Algorithm \ref{alg:overall-section4} is given by $\mathcal{O}(I_{\mathrm{AO}}\max\{2N^3,2N^2Q, M^3, Q^3,\mathcal{C}_{FW}\})$, where $I_{\mathrm{AO}}$ denotes the number of outer iterations.

	\vspace{-3mm}
	\section{Extension to Multi-Receiver Case}  \label{Extension to The Multi-Receivers Case}
	\vspace{-2mm}
	In this section, we extend the proposed framework to a multicast secure transmission scenario with \(K\ge 2\) legitimate receivers, where all legitimate receivers intend to decode the same confidential message. 
	Due to the worst-receiver SR objective, the resulting optimization problem involves a challenging min-max structure. 
	Therefore, instead of directly solving the original multicast SR maximization problem, we derive a tractable lower bound and optimize this lower-bound objective via an AO-based algorithm.
	Let \(\mathcal K=\{1,2,\ldots,K\}\) denote the set of legitimate receivers. For the $k$-th legitimate receiver, the local boresight direction of its $m$-th receive RA is denoted by $\mathbf f_{r,k,m}\in\mathbb R^3$. 
	Then, the collection of receive RA orientations at receiver $k$ is defined~as
	\begin{align}
		\mathbf f_{r,k}=\left[\mathbf f_{r,k,1},\mathbf f_{r,k,2},\ldots,\mathbf f_{r,k,M}\right]\in\mathbb R^{3\times M},  \forall k \in  \mathcal{K}.
	\end{align}
	The signal model for the MIMO multi-receiver wiretap channel case is given by
	\begin{align}
		\mathbf{y}_k = \mathbf{H}_k (\mathbf{W} \mathbf{s} + \mathbf{z})+\mathbf{n}_k,  \forall k \in  \mathcal{K},
	\end{align}
	where $\mathbf{H}_k$  denotes the channel from the transmitter to the $k$-th receiver. Then, the achievable rate of receiver $k$ is
	\begin{align}
		&R_k \!= \! \log_2\! \det  \left( \mathbf{I}_{M} \!+\! \mathbf{H}_k \mathbf{W} \mathbf{W}^H \mathbf{H}_k^H  \bm{\Sigma}_k^{-1} \right),  \forall k \in  \mathcal{K},
	\end{align}
	where the interference-plus-noise covariance matrix at the $k$-th receiver is given by $\bm{\Sigma}_k = \mathbf{H}_k \mathbf{W}_e\mathbf{W}_e^H \mathbf{H}_k^H + \sigma_k^2 \mathbf{I}_{M}$.
	The SR is formulated as  $R_{\mathrm{s}} =\min_{k \in \mathcal{K}}  [{R}_k -  R_{e}]_{+}$.
	The related problem in the multicast scenario is expressed as 
	\begin{subequations}
		\label{P4}
		\begin{eqnarray}
			\label{P4-0}
			&\!\!\!\!\!\!\!\!\!\!\!\!\!\!\!\!\! \max  \limits_{\mathbf{f}_{t},\mathbf{f}_{r,k}, \mathbf{W},\mathbf{W}_e}  
			&\!\!\!\!\!\! R_{\mathrm{s}}   \\
			\label{P4-1}
			&\!\!\!\!\!\!\!\!\!\!\!\!\!\!\!\!\! \mathrm{s.t.}  &\!\!\!\!\!\! \mathrm{Tr}(\mathbf{W}\mathbf{W}^H+\mathbf{W}_e\mathbf{W}_e^H) \leq P_{\max}, \\
			\label{P4-2}
			&&\!\!\!\!\!\! \cos(\theta_{\max}) \le \mathbf{f}_{t,n}^T\mathbf{e}_z \le 1,  \forall n \in \mathcal{N},\\
			\label{P4-3}
			&&\!\!\!\!\!\! \cos(\theta_{\max}) \le \mathbf{f}_{r,k,m}^T\mathbf{e}_z \le 1, \forall m \in \mathcal{M},\\
			\label{P4-4}
			&&\!\!\!\!\!\! \Vert \mathbf{f}_{t,n} \Vert_2 \!\!=\!\! 1, \Vert \mathbf{f}_{r,k,m} \Vert_2 \!\!=\!\! 1, \forall n \in \mathcal{N}, \forall m \in \mathcal{M}.
		\end{eqnarray}
	\end{subequations} 
	Similar to the previous processing, we rewrite the objective function \eqref{P4-0} as
	\begin{align}
		\label{Mul-SR}
		\!\!\! R_{\mathrm{s}} \!=&\! \min_{k \in \mathcal{K}} \left\{ \log_2 \det  \left( \mathbf{I}_{M} \!+\! \mathbf{H}_k \mathbf{W} \mathbf{W}^H \mathbf{H}_k^H  \bm{\Sigma}_k^{-1} \right) \right\} \nonumber  \\
		& \! -\log_2 \! \det  \left( \mathbf{I}_{Q} \!+\! \mathbf{H}_e \mathbf{W} \mathbf{W}^H \mathbf{H}_e^H \bm{\Sigma}_e^{-1} \right)   \nonumber \\
		\!=&\!\min_{k \in \mathcal{K}} \left\{ \log_2 \det  \left( \mathbf{I}_{M} + \mathbf{H}_k \mathbf{W} \mathbf{W}^H \mathbf{H}_k^H  \bm{\Sigma}_k^{-1} \right) \right\} \nonumber \\
		&\!+\log_2 \! \det  \left( \mathbf{I}_{Q} \!+\! \mathbf{H}_e \mathbf{W}_e \mathbf{W}_e^H \mathbf{H}_e^H (\sigma_e^2 \mathbf{I}_{Q})^{-1} \right)   \nonumber \\
		& \!-\log_2 \! \det  \left( \mathbf{I}_{Q} \!+\! \sigma_e^{-2}\mathbf{H}_e (\mathbf{W} \mathbf{W}^H + \mathbf{W}_e \mathbf{W}_e^H) \mathbf{H}_e^H  \right)  \nonumber \\
		& = {1}/{\mathrm{ln2}} \Xi,\\
		\!\!\! \Xi \!=&\!\min_{k \in \mathcal{K}} \left\{ \max_{\{\mathbf{U}_k, \mathbf{\Omega}_k\}_{k=1}^K} h_{1,k} (\mathbf{U}_k, \mathbf{\Omega}_k, \mathbf{W},\mathbf{W}_e) \right\}  \nonumber \\
		&\!+\!\! \max_{\mathbf{U}_{e}, \mathbf{\Omega}_{e}} h_{2} (\mathbf{U}_{e}, \mathbf{\Omega}_{e}, \mathbf{W}_e) \!\!+\!\! \max_{\mathbf{\Omega}_{x}} h_{3} (\mathbf{\Omega}_{x}, \mathbf{W},\mathbf{W}_e),
	\end{align}
	The first term in \eqref{Mul-SR} involves a nested min-max structure, i.e., $\min_{k\in\mathcal{K}} \max_{\{ \mathbf{U}_k,\boldsymbol{\Omega}_k\}} h_{1,k}$, which is difficult to handle directly. In particular, the outer minimization over receivers is coupled with the inner maximization over the auxiliary variables, making it intractable to derive simple block updates as in the single-receiver case. To obtain a tractable reformulation, we derive a lower bound for the first term of $(\ref{Mul-SR})$
	\begin{align}
		\label{Mul-SR1}
		&\min_{k \in \mathcal{K}} \left\{ \max_{\{\mathbf{U}_k, \mathbf{\Omega}_k\}_{k=1}^K} h_{1,k} (\mathbf{U}_k, \mathbf{\Omega}_k, \mathbf{W},\mathbf{W}_e) \right\} \nonumber \\
		& \ge \max_{\{\mathbf{U}_k, \mathbf{\Omega}_k\}_{k=1}^K} \left\{  \min_{k \in \mathcal{K}} h_{1,k} (\mathbf{U}_k, \mathbf{\Omega}_k, \mathbf{W},\mathbf{W}_e) \right\},
	\end{align}
	where $\min_x \max_y f(x,y) \ge \max_y \min_x f(x,y)$ holds for any $f(x,y)$.
	Then, by substituting the first term of $(\ref{Mul-SR})$ with the right-hand side of $(\ref{Mul-SR1})$, we maximize the lower bound of the original SR as
	\begin{subequations}
		\label{P5}
		\begin{eqnarray}
			\label{P5-0}
			&\!\!\!\!\!\!\!\!\!\!\!\!\!\!\!\!\! \max  \limits_{\mathbf{K}_m,\mathbf{W}_e, \mathbf{W}, \mathbf{f}_t, \mathbf{f}_{r,k}}  
			&\!\!\!\!\!\! \bar{F} (\mathbf{K}_m,\mathbf{W}_e, \mathbf{W}, \mathbf{f}_t, \mathbf{f}_{r,k}) \\
			\label{P5-1}
			&\!\!\!\!\!\!\!\!\!\! \mathrm{s.t.}  &\!\!\!\!\!\!
			\mathbf{\Omega}_k \succeq \mathbf{0},\mathbf{\Omega}_e \succeq \mathbf{0},\mathbf{\Omega}_x \succeq \mathbf{0},\\
			\label{P5-2}
			&&\!\!\!\!\!\! (\mathrm{\ref{P4-1}}),(\mathrm{\ref{P4-2}}),(\mathrm{\ref{P4-3}}),(\mathrm{\ref{P4-4}}),
		\end{eqnarray}
	\end{subequations}
	where $\mathbf{K}_m \triangleq [\mathbf{U}_k,\mathbf{\Omega}_k,\mathbf{U}_{e}, \mathbf{\Omega}_{e},\mathbf{\Omega}_{x}] $ and $\bar{F} = \min_{k \in \mathcal{K}} h_{1,k} (\mathbf{U}_k, \mathbf{\Omega}_k, \mathbf{W},\mathbf{W}_e) + h_{2} (\mathbf{U}_{e}, \mathbf{\Omega}_{e}, \mathbf{W}_e) + h_{3} ( \mathbf{\Omega}_{x}, \mathbf{W},\mathbf{W}_e)$. The terms $h_{2} (\mathbf{U}_{e}, \mathbf{\Omega}_{e}, \mathbf{W}_e)$ and $h_{3} ( \mathbf{\Omega}_{x}, \mathbf{W},\mathbf{W}_e)$ are given in \eqref{h2} and \eqref{h3}. We first update all auxiliary matrices $\{\mathbf{U}_k,\mathbf{\Omega}_k,\mathbf{U}_{e}, \mathbf{\Omega}_{e},\mathbf{\Omega}_{x}\}$ in closed forms with fixed $\{\mathbf{W}_e, \mathbf{W}, \mathbf{f}_t, \mathbf{f}_{r,k}\}$ following the same procedure as in Section \ref{Proposed Solution}. Then, we optimize $\{\mathbf{W}_e, \mathbf{W}, \mathbf{f}_t, \mathbf{f}_{r,k}\}$ by solving the following problem
	\begin{subequations}
		\label{P6}
		\begin{eqnarray}
			\label{P6-0}
			&\!\!\!\!\!\!\!\!\!\!\!\!\!\!\!\!\! \min  \limits_{\mathbf{W}_e, \mathbf{W}, \mathbf{f}_t, \mathbf{f}_{r,k}}  
			&\!\!\!\!\!\! \bar{L}(\mathbf{W}_e, \mathbf{W}, \mathbf{f}_t, \mathbf{f}_{r,k}) \\
			\label{P6-1}
			&\!\!\!\!\!\!\!\!\!\! \mathrm{s.t.}  &\!\!\!\!\!\! (\mathrm{\ref{P4-1}}),(\mathrm{\ref{P4-2}}),(\mathrm{\ref{P4-3}}),(\mathrm{\ref{P4-4}}),
		\end{eqnarray}
	\end{subequations}
	where 
	\begin{align}
		&\bar{L}(\mathbf{W}_e, \mathbf{W}, \mathbf{f}_t, \mathbf{f}_{r,k})
		= \nonumber \\
		& \max_{k \in \mathcal{K}} \left\{ -2 \Re \{\mathrm{Tr}(\mathbf{\Omega}_k \mathbf{W} \mathbf{H}_k^H \mathbf{U}_k)\} + \mathrm{Tr}\left(\mathbf{W}^H  \overline{\mathbf{H}}_k  \mathbf{W} \right) \right. \nonumber \\
		&\left. + \mathrm{Tr}\left(\mathbf{W}_e^H  \overline{\mathbf{H}}_k  \mathbf{W}_e \right) - C_k \right\}+\mathrm{Tr}\left(\mathbf{W}^H  \overline{\mathbf{H}}_e^x  \mathbf{W} \right)\nonumber \\
		&  - 2 \Re\left\{ \mathrm{Tr}\left( \mathbf{\Omega}_e \mathbf{W}_e^H {\mathbf{H}}_e^H \mathbf{U}_e \right) \right\} +\mathrm{Tr}\left(\mathbf{W}_e^H  \overline{\mathbf{H}}_e  \mathbf{W}_e \right),\\
		& \overline{\mathbf{H}}_k = \mathbf{H}_k^H\mathbf{U}_k\mathbf{\Omega}_k\mathbf{U}_k^H\mathbf{H}_k, \overline{\mathbf{H}}_e^x = \sigma_e^{-2}\mathbf{H}_e^H\mathbf{\Omega}_x\mathbf{H}_e, \\
		&\overline{\mathbf{H}}_e = \mathbf{H}_e^H\mathbf{U}_e\mathbf{\Omega}_e\mathbf{U}_e^H\mathbf{H}_e+ \sigma_e^{-2}\mathbf{H}_e^H\mathbf{\Omega}_x\mathbf{H}_e,\\
		&C_k = \ln \mathrm{det}(\mathbf{\Omega}_k) +d - \mathbf{Tr} \left(\mathbf{\Omega}_k + \sigma_k^{2}\mathbf{\Omega}_k\mathbf{U}_k^H\mathbf{U}_k \right).
	\end{align}

	\vspace{-6mm}
	\subsection{Transmit Beamforming Design}
	\vspace{-1mm}
	With fixed $\mathbf{f}_t$ and $\mathbf{f}_{r,k}$, we optimize $\mathbf{W}$ and $\mathbf{W}_e$. The subproblem is given by
		\begin{eqnarray}
			\label{P7-0}
			& \min  \limits_{\mathbf{W}_e, \mathbf{W}}  
			\bar{L}(\mathbf{W}_e, \mathbf{W}) \ \ \ \mathrm{s.t.}  & \eqref{P4-1},
		\end{eqnarray}
	which is a convex QCQP problem with a min-max form. We first introduce a slack variable $\vartheta$, then we have
	\begin{subequations}
		\label{P8}
		\begin{eqnarray}
			\label{P8-0}
			&\!\!\!\!\!\!\!\!\!\!\!\!\!\!\!\!\! \min  \limits_{\mathbf{W}_e, \mathbf{W},\vartheta}  
			&\!\!\!\! \hat{L}(\mathbf{W}_e, \mathbf{W}) \\
			\label{P8-1}
			&\!\!\!\!\!\!\!\!\!\! \mathrm{s.t.}  &\!\!\!\! \mathrm{Tr}(\mathbf{W}\mathbf{W}^H+\mathbf{W}_e\mathbf{W}_e^H) \leq P_{\max},\\
			&&\!\!\!\! -2 \Re \{\mathrm{Tr}(\mathbf{\Omega}_k \mathbf{W} \mathbf{H}_k^H \mathbf{U}_k)\} + \mathrm{Tr}\left(\mathbf{W}^H  \overline{\mathbf{H}}_k  \mathbf{W} \right)  \nonumber \\
			&&\!\!\!\! + \mathrm{Tr}\left(\mathbf{W}_e^H  \overline{\mathbf{H}}_k  \mathbf{W}_e \right) - C_k \le \vartheta, \forall k,
		\end{eqnarray}
	\end{subequations}
	where 
	\begin{align}
		\hat{L} = &\vartheta +\mathrm{Tr}\left(\mathbf{W}^H  \overline{\mathbf{H}}_e^x  \mathbf{W} \right) - 2 \Re\left\{ \mathrm{Tr}\left( \mathbf{\Omega}_e \mathbf{W}_e^H {\mathbf{H}}_e^H \mathbf{U}_e \right) \right\} \nonumber \\
		& +\mathrm{Tr}\left(\mathbf{W}_e^H  \overline{\mathbf{H}}_e  \mathbf{W}_e \right),
	\end{align}
	Thus, problem (\ref{P8}) can be solved by CVX~\cite{cvx}.
	
	\vspace{-4mm}
	\subsection{RA Orientation Design}
	\vspace{-1mm}
	Fixed \{$\mathbf{W}$, $\mathbf{W}_e$\}, we optimize the transmit/receive RA orientations $\mathbf{f}_t$ and $ \mathbf{f}_{r,k}$. The related subproblem is given by
		\begin{eqnarray}
			\label{P9-0}
			& \min  \limits_{\mathbf{f}_t, \mathbf{f}_{r,k}}  
			& \bar{L}(\mathbf{f}_t, \mathbf{f}_{r,k}) 
			\ \ \  \mathrm{s.t.}  (\mathrm{\ref{P4-2}}),(\mathrm{\ref{P4-3}}),(\mathrm{\ref{P4-4}}).
		\end{eqnarray}
	For the objective function in problem $(\mathrm{\ref{P9-0}})$, we have
	\begin{align}
		\!\!\!\!\!\! \hat{L} =& \vartheta +\mathrm{Tr}\left(\mathbf{W}^H  \overline{\mathbf{H}}_e^x  \mathbf{W} \right) - 2 \Re\left\{ \mathrm{Tr}\left( \mathbf{\Omega}_e \mathbf{W}_e^H {\mathbf{H}}_e^H \mathbf{U}_e \right) \right\} \nonumber \\
		& +\mathrm{Tr}\left(\mathbf{W}_e^H  \overline{\mathbf{H}}_e  \mathbf{W}_e \right),\nonumber \\
		=&\vartheta - 2 \Re\left\{ \mathrm{Tr}\left(   {\mathbf{D}}_e\mathbf{H}_e^H \right) \right\} \nonumber \\
		&+\mathrm{Tr}\left( \mathbf{H}_e \mathbf{W}_X  \mathbf{H}_e^H \mathbf{C}_X\right)
		\!+\!\mathrm{Tr}\left(\mathbf{H}_e\mathbf{W}_e\mathbf{W}_e^H   \mathbf{H}_e^H\mathbf{C}_e   \right),
	\end{align}
	where $\mathbf{W}_X = \mathbf{W}\mathbf{W}^H + \mathbf{W}_e\mathbf{W}_e^H $,   $\mathbf{C}_X=\sigma_e^{-2}\mathbf{\Omega}_x$,  $\mathbf{C}_e=\mathbf{U}_e\mathbf{\Omega}_e\mathbf{U}_e^H$, and $\mathbf{D}_e=\mathbf{U}_e\mathbf{\Omega}_e \mathbf{W}_e^H$. 
	Then, we have
	\begin{subequations}
		\label{P10}
		\begin{eqnarray}
			\label{P10-0}
			&\!\!\!\!\!\!\!\!\!\!\!\!\!\!\!\!\!\!\!\!\! \min  \limits_{\mathbf{f}_t, \mathbf{f}_{r,k}}  
			&\!\!\!\!\!\! \hat{L}(\mathbf{f}_t, \mathbf{f}_{r,k}) \\
			\label{P10-1}
			&\!\!\!\!\!\!\!\!\!\!\!\!\!\!\!\!\!\!\!\!\! \mathrm{s.t.}  &\!\!\!\!\!\!  g_k(\mathbf{f}_t,\mathbf{f}_{r,k}) \le \vartheta, \forall k\in\mathcal{K}, (\mathrm{\ref{P4-2}}),(\mathrm{\ref{P4-3}}),(\mathrm{\ref{P4-4}}),
		\end{eqnarray}
	\end{subequations}
	where 
	\begin{align}
		&\!\! g_k(\mathbf{f}_t,\mathbf{f}_{r,k}) \!=\! \mathrm{Tr}\left(\mathbf{H}_k\mathbf{W}_X  \mathbf{H}_k^H\mathbf{C}_k   \right) 
		\!\!-\!\! 2 \Re \{\mathrm{Tr}(\mathbf{D}_k \mathbf{H}_k^H )\} \!\!-\!\! C_k, \nonumber \\
		&\!\! \mathbf{C}_k=\mathbf{U}_k\mathbf{\Omega}_k\mathbf{U}_k^H, \mathbf{D}_k=\mathbf{U}_k\mathbf{\Omega}_k \mathbf{W}^H. \nonumber
	\end{align}
	The Lagrange function is constructed as follows
	\begin{align}
		\label{eqObjective-function}
		&L(\mathbf{f}_t, \mathbf{f}_r) \!=\!\vartheta \!-\! 2 \Re\left\{ \mathrm{Tr}\left(   {\mathbf{D}}_e\mathbf{H}_e^H \right) \right\} \!+\! \mathrm{Tr}\left( \mathbf{H}_e \mathbf{W}_X  \mathbf{H}_e^H \mathbf{C}_X\right) \nonumber \\
		&
		\!+\!\mathrm{Tr}\left(\mathbf{H}_e\mathbf{W}_e\mathbf{W}_e^H\mathbf{H}_e^H\mathbf{C}_e \right)\!+\! \sum\limits_{k=1}^K \mu_k(g_k(\mathbf{f}_t,\mathbf{f}_{r,k}) \!-\! \vartheta),
	\end{align}
	where $\mu_k\ge 0$ is the Lagrange multipliers with
	\begin{align}
		& \!\!\!\! \mu_k^{(t+1)} 
		\!=\! \Big[\mu_k^{(t)} \!+\! \beta^{(t)}\big(g_k(\mathbf{f}_t^{(t+1)},\mathbf{f}_{r,k}^{(t+1)})\!-\!\vartheta^{(t+1)}\big)\Big]_+, \forall k.   \label{eq:mu_update}
	\end{align}
	The term $\beta^{(t)}>0$ is an outer-loop stepsize. $\mathbf{f}_t^{(t+1)}$ and $\mathbf{f}_{r,k}^{(t+1)}$ are the values of $\mathbf{f}_t$ and $\mathbf{f}_{r,k}$ in the $(t+1)$-th iteration.
	The variable $\vartheta^{(t+1)}$ can be updated in the $(t+1)$-th iteration by 
	\begin{align}
		&
		\vartheta^{(t+1)}=\max_{k \in \mathcal{K}}g_k(\mathbf{f}_t^{(t+1)},\mathbf{f}_{r,k}^{(t+1)}).
		\label{eq:vartheta_update}
	\end{align}
	By \eqref{eqObjective-function}, we have the Wirtinger derivatives
	\begin{align}
		&\!\!\! \mathbf{G}_{e} \triangleq \frac{\partial L}{\partial \mathbf{H}_e^{*}} = \mathbf{C}_X \mathbf{H}_e \mathbf{W}_X
		+ \mathbf{C}_e \mathbf{H}_e(\mathbf{W}_e\mathbf{W}_e^H)
		- \mathbf{D}_e, \\
		&\!\!\! \mathbf{G}_{k} \triangleq \frac{\partial L}{\partial \mathbf{H}_k^{*}}=\mu_k\big(\mathbf{C}_k\mathbf{H}_k\mathbf{W}_X-\mathbf{D}_k\big), \forall k.
	\end{align}
	Let $[\mathbf{H}_k]_{m,n}=h_{k,m,n}$ and $[\mathbf{H}_e]_{q,n}=h_{e,q,n}$.
	For each transmit RA orientation $\mathbf{f}_{t,n}\in\mathbb{R}^3$, we have
	\begin{align}
		\nabla_{\mathbf{f}_{t,n}} L
		=&
		\sum\limits_{k=1}^K \sum\limits_{m=1}^M
		2\,\Re\!\left\{
		[\mathbf{G}_k]_{m,n}^{\ast}\,
		\frac{\partial h_{k,m,n}}{\partial \mathbf{f}_{t,n}}
		\right\} \nonumber \\
		&+
		\sum\limits_{q=1}^Q
		2\,\Re\!\left\{
		[\mathbf{G}_e]_{q,n}^{\ast}\,
		\frac{\partial h_{e,q,n}}{\partial \mathbf{f}_{t,n}}
		\right\}.
		\label{eq:grad_ft}
	\end{align}
	For each receive RA orientation $\mathbf{f}_{r,k,m}\in\mathbb{R}^3$, we get
	\begin{align}
		\nabla_{\mathbf{f}_{r,k,m}} L
		=
		\sum\limits_{n=1}^N
		2\,\Re\!\left\{
		[\mathbf{G}_k]_{m,n}^{\ast}\,
		\frac{\partial h_{k,m,n}}{\partial \mathbf{f}_{r,k,m}}
		\right\},
		\forall k, \forall m.
		\label{eq:grad_fr}
	\end{align}
	The terms $\frac{\partial h_{k,m,n}}{\partial \mathbf{f}_{t,n}}$, $\frac{\partial h_{k,m,n}}{\partial \mathbf{f}_{r,k,m}}$, and $\frac{\partial h_{e,q,n}}{\partial \mathbf{f}_{t,n}}$ can be obtained by the same procedure in Appendix \ref{Appendix B}, which is omitted for simplicity. Regarding the computational complexity, let \(I_{\rm AO}\), \(I_{\rm W}\), \(I_{\rm FW}\), and \(I_A\) denote the numbers of AO iterations, interior-point iterations, Riemannian Frank-Wolfe iterations, and Armijo backtracking iterations, respectively.
	The complexity of updating the auxiliary variables \(\{\mathbf{U}_k,\bm{\Omega}_k\}_{k=1}^K\), \(\mathbf{U}_e\), \(\bm{\Omega}_e\), and \(\bm{\Omega}_x\), is given by $\mathcal{C}_{\rm aux}=\mathcal{O}\left(K(M^3+d^3)+N^3+2Q^3\right)$. The complexity of computing the transmit beamforming $\mathbf{W}$ and $\mathbf{W}_e$ is given by $\mathcal{C}_{\rm W}=\mathcal{O}\left(I_{\rm W}\left[(2Nd+2N^2+1)^3+(K+1)(2Nd+2N^2+1)^2	\right]\right)$. Moreover, the complexity of the RA orientation update is $\mathcal{C}_{\rm FW}=\mathcal{O}\left(I_{\rm FW}\left[2KMN+NQ+I_A(KMN+NQ)\right]\right)$. Therefore, the overall computational complexity of the proposed algorithm for the multi-receiver case is $\mathcal{C}_{\rm total}=\mathcal{O}\left(I_{\rm AO}\left[\mathcal{C}_{\rm aux}+\mathcal{C}_{\rm W}+\mathcal{C}_{\rm FW}\right]\right)$.

	\vspace{-4mm}
	\section{Simulation Results}  \label{Simulation Results}
	\vspace{-1mm}
	In this section, simulation results are provided to verify the effectiveness of the proposed algorithm and to illustrate the secrecy performance benefits of RA-enabled secure MIMO transmission.

	\vspace{-4mm}
	\subsection{Simulation Setup}
	\vspace{-2mm}
	In simulation, we assume the system operates at a frequency of 3.5 GHz, with a wavelength of $\lambda \!=\! 0.0857$ m. The antenna separation at the transmitter and receiver is set to $\lambda/2$. The transmitter and the legitimate receiver are equipped with $N \!=\!25$ and $M \!= \! 4$ RAs, respectively, and the eavesdropper has $Q \!= \! 4$ fixed antennas. The number of data streams is $d=2$. Besides, the distance from the transmitter to each legitimate receiver is independently generated as 
	\(d_k\sim\mathcal U(d_{\min},d_{\max})\), \(k\in\mathcal K\), 
	while the distance from the transmitter to the eavesdropper is independently generated as 
	\(d_e\sim\mathcal U(d_{\min},d_{\max})\).
	Unless otherwise specified, \(d_{\min}=20\) m and \(d_{\max}=40\) m. We assume the maximum transmit power of $P_{\max}= 20 ~\mathrm{dBm}$. The noise powers are set as \(\sigma_0^2=\sigma_k^2=\sigma_e^2=-80\) dBm.
	We consider the following baselines.
	\begin{itemize}
		\item \textbf{RFOA}: In this scheme, antennas at the receiver are fixed.
		\item \textbf{FOA}: In this scheme, antennas at both the transmitter and the receiver are fixed.
		\item \textbf{Isotropic-antenna baseline:} In this scheme, the directional gain is set to $p=0$.
		\item \textbf{Random-orientation baseline}: In this scheme, the deflection angles of each RA are randomly generated, satisfying the maximum zenith angle constraint \eqref{theta_max}.
		\item \textbf{Discrete-orientation baseline}: The orientation of each antenna is restricted to a discrete set. To avoid the uniform grids, we adopt spherical Fibonacci sampling, which generates nearly uniform and equal-area reference directions over the cap. The $i$-th samples are given by
		$\theta_{i} = \arccos \left( 1 - \frac{i + \frac{1}{2}}{N_{\text{sam}}} (1 - \cos \theta_{\max}) \right),
		\phi_{i} = 2\pi \, \text{frac} \left( \frac{i}{\varphi^{2}} \right), \forall i$, where $N_{\mathrm{sam}} = 32$ is the codebook size, $\varphi = (1+\sqrt{5})/2$ is the golden ratio, and $\mathrm{frac}(x)=x-\lfloor x \rfloor$ denotes the fractional part \cite{Peng2025}.
		The nearest projection scheme is applied to obtain the final orientations.
	\end{itemize}
	
	\vspace{-4mm}
	\subsection{Convergence Behavior of the proposed Algorithm}
	\vspace{-1mm}
	\begin{figure}[t]
		\centering
		\includegraphics[width=0.4 \textwidth]{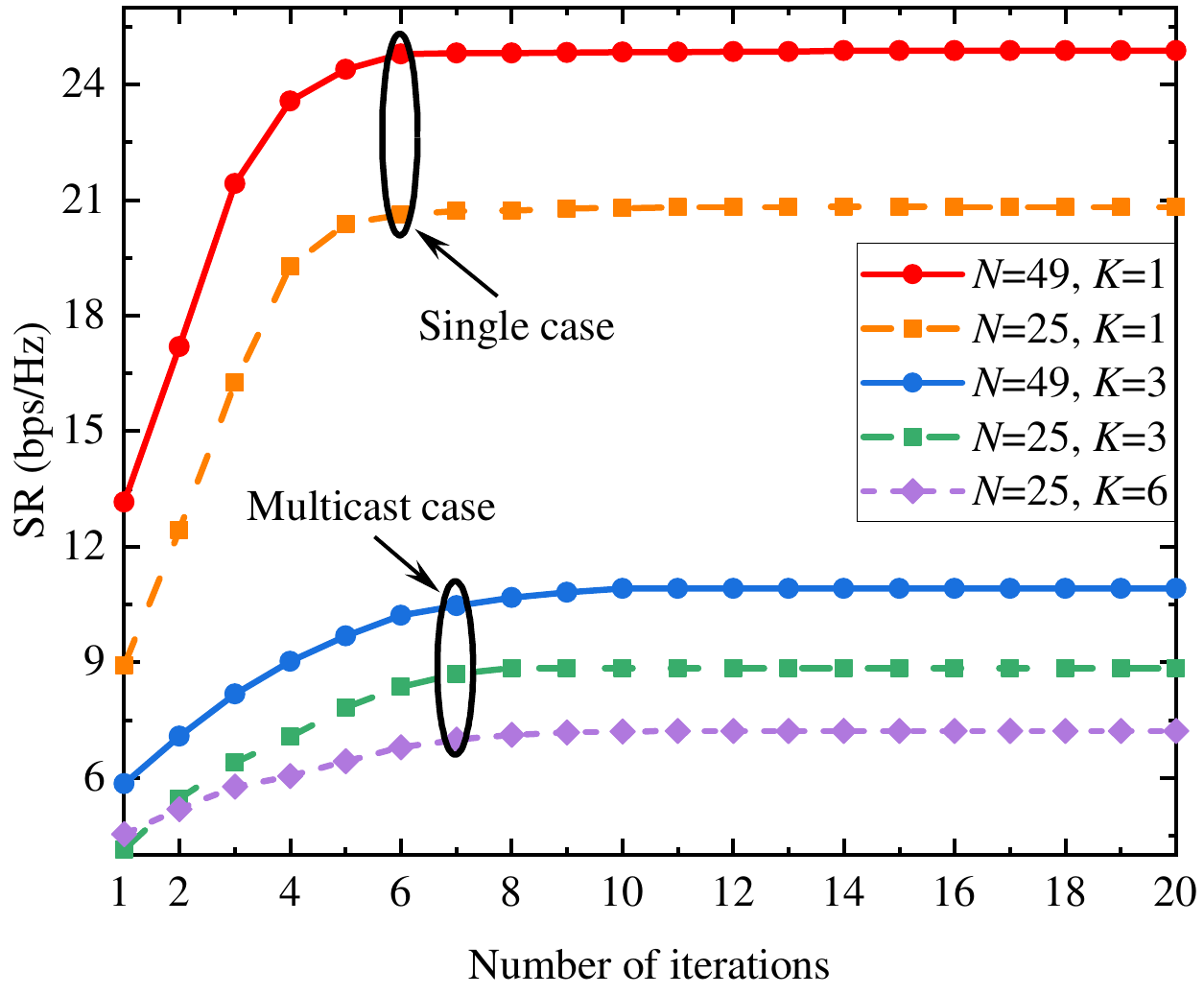}
		\caption{Convergence behavior of the proposed algorithm under single-receiver and multicast secure transmission scenarios.}
		\label{Fig2}
		\vspace{-5mm}
	\end{figure}
	Fig.~\ref{Fig2} illustrates the convergence behavior of the proposed algorithm under different numbers of $N$ and $K$. As observed, for all considered settings, the proposed algorithm converges rapidly and stably, and the SR reaches a nearly steady value within approximately $10$ iterations, thereby verifying its effectiveness and computational efficiency. In addition, the convergence becomes relatively slower as $K$ increases. The reason is that a larger number of receivers introduces more coupled constraints and leads to a more complicated optimization problem, which generally requires more iterations to approach a stationary solution. Meanwhile, the SR decreases with increasing $K$, which is intuitive because the multicast secure transmission rate is constrained by the worst-case receiver, and accommodating more receivers makes it more difficult to simultaneously guarantee favorable legitimate links while suppressing information leakage. Furthermore, for larger $N$, the algorithm converges to a higher SR, indicating that more transmit antennas provide additional spatial DoFs for beamforming and secure transmission design.
	
	\vspace{-4mm}
	\subsection{Performance Analysis}
	\vspace{-1mm}
	\begin{figure}[t]
		\centering
		\includegraphics[width=0.4 \textwidth]{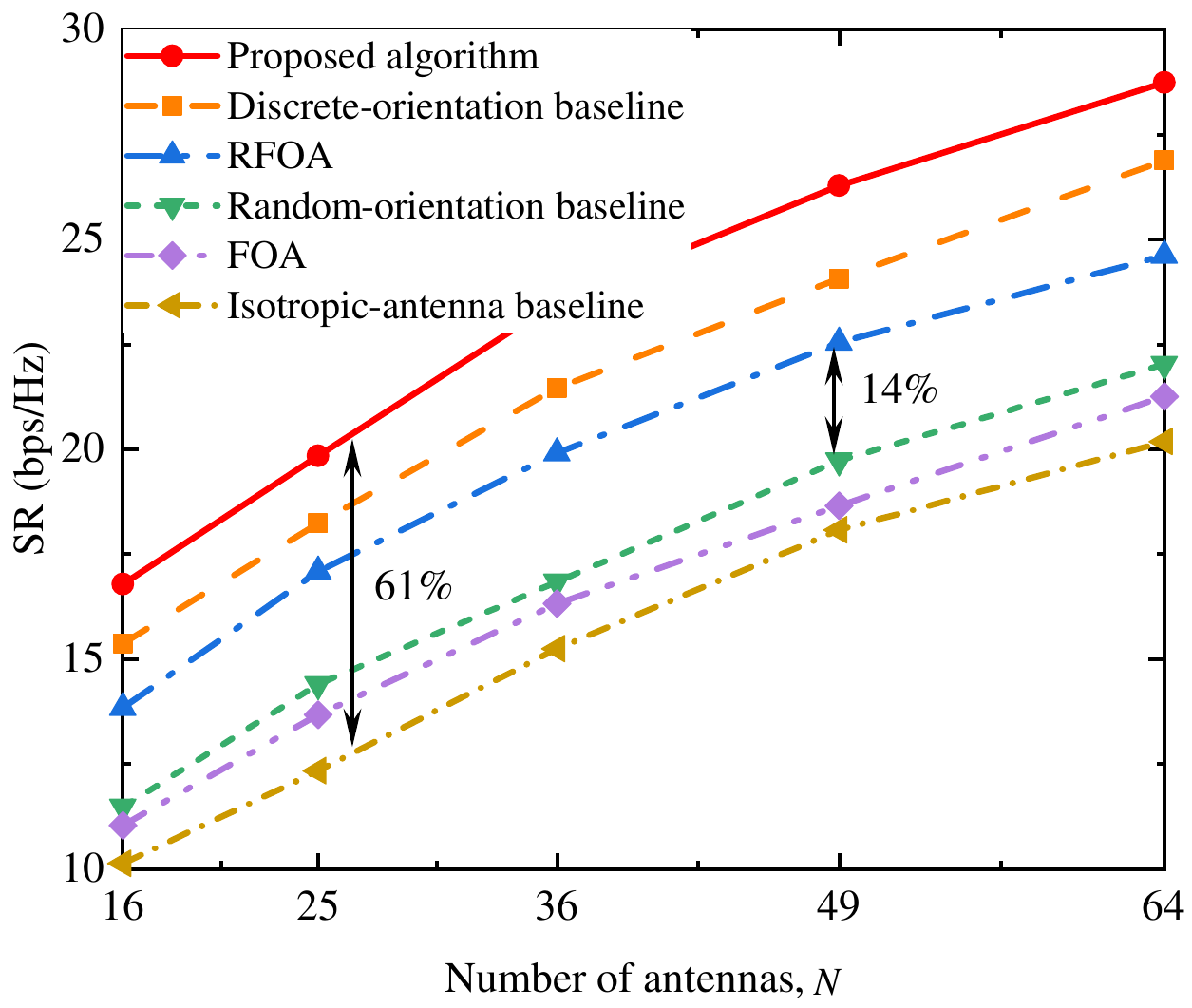}
		\caption{SR versus the number of transmit antennas.}
		\label{Fig3}
		\vspace{-7mm}
	\end{figure}
	
	Fig.~\ref{Fig3} depicts the SR versus the number of transmit antennas under different schemes. The SR of all schemes generally improves as $N$ increases, since more transmit antennas offer additional spatial DoFs for enhancing legitimate transmission and suppressing information leakage through beamforming and AN design.
	Among all considered schemes, the proposed algorithm consistently achieves the highest SR over the entire range, which demonstrates the effectiveness of the proposed algorithm. Specifically, when $N = 25$, the proposed algorithm achieves about $61\%$ higher SR than the isotropic-antenna baseline. Moreover, since the orientation of the transmit antennas can be optimized, the RFOA scheme performs better than the random-orientation baseline, where it enjoys a $14\%$ performance gain when $N=49$.
	
	\begin{figure}[t]
		\centering
		\includegraphics[width=0.4 \textwidth]{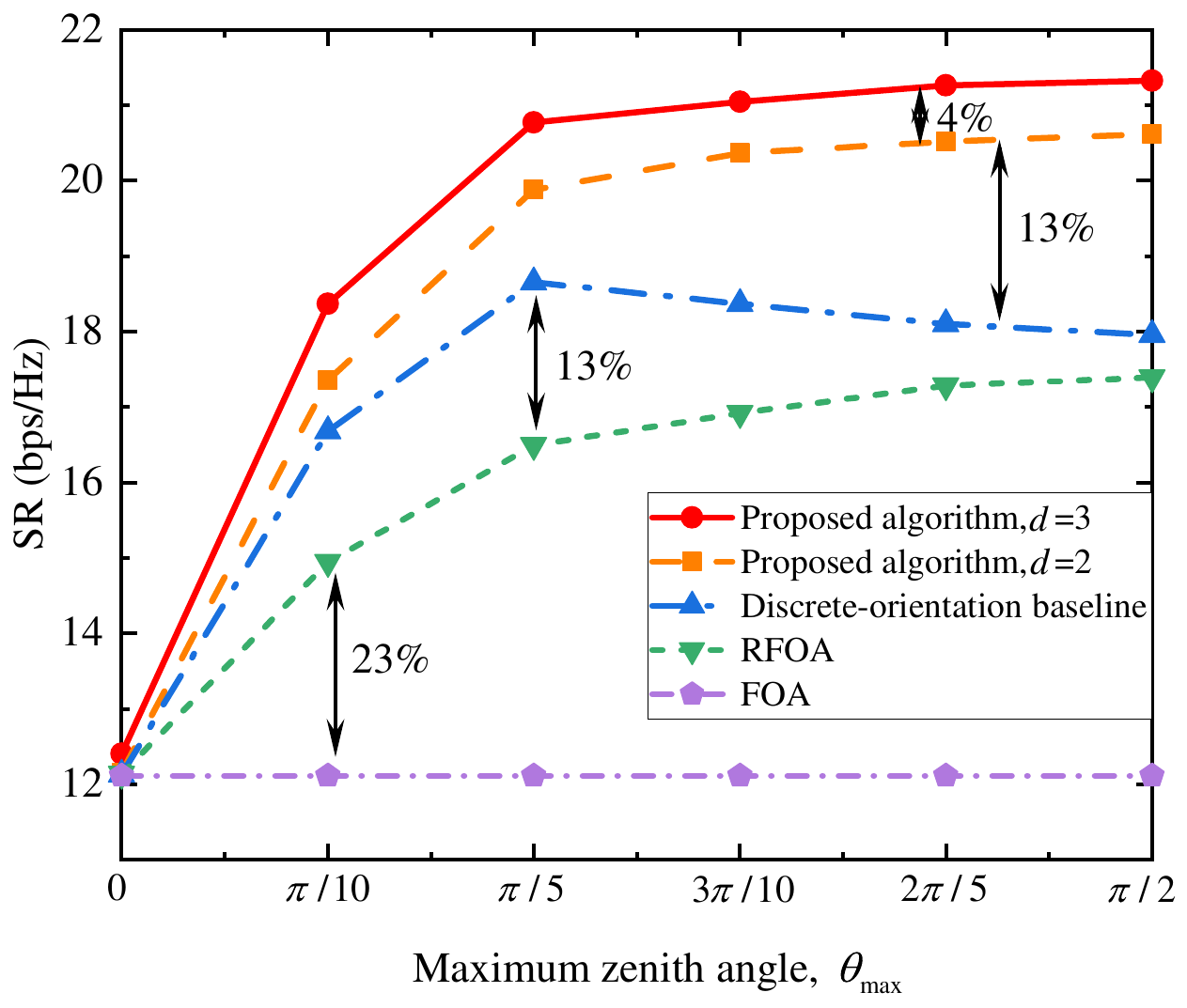}
		\caption{SR versus the maximum zenith angle.}
		\label{Fig4}
		\vspace{-8mm}
	\end{figure}
	
	Fig.~\ref{Fig4} shows the SR versus the maximum zenith angle under different schemes. It can be observed that from $0$ to $\pi/5$, the SR of the proposed algorithm increases significantly and consistently outperforms the benchmarks, indicating that a small zenith angle change provides a higher achievable secrecy performance improvement. After $\pi/5$, the SR improvement gradually becomes marginal, especially for the proposed algorithm, which suggests that the gain brought by increasing the angular range becomes saturated in the later zenith angle region. In particular, when $d=3$, the proposed algorithm achieves the best performance over the entire  region, achieving a $4\%$ performance gain over the case with $d=2$. The case with $d=2$ maintains a clear advantage over the discrete-orientation baseline with a $13\%$ performance gain.
	Moreover, the discrete-orientation baseline first increases from $0$ to $\pi/5$, then decreases after $\pi/5$. The main reason is when the maximum zenith angle increases from $0$ to $\pi/5$, the enlarged angular range provides more flexibility for antenna rotation, such that the selected discrete angle can better approach the continuous optimal solution. However, with the increase of the $\theta_{\mathrm{max}}$, the mismatch between the continuous optimal angle and discrete projection becomes more pronounced. In this case, the quantization loss gradually outweighs the benefit brought by the enlarged angular range. Consequently, the selected discrete directions cannot effectively enhance the legitimate links and suppress information leakage simultaneously, which leads to a gradual decrease in SR. Specifically, at $\theta_{\max}=\pi/5$, the discrete-orientation baseline achieves about $13\%$ higher SR than the RFOA baseline.
	In addition, at $\theta_{\max}=\pi/10$, the RFOA baseline outperforms the FOA baseline by about $23\%$, demonstrating the benefits of antenna rotation.

	\begin{figure}[t]
		\centering
		\includegraphics[width=0.4 \textwidth]{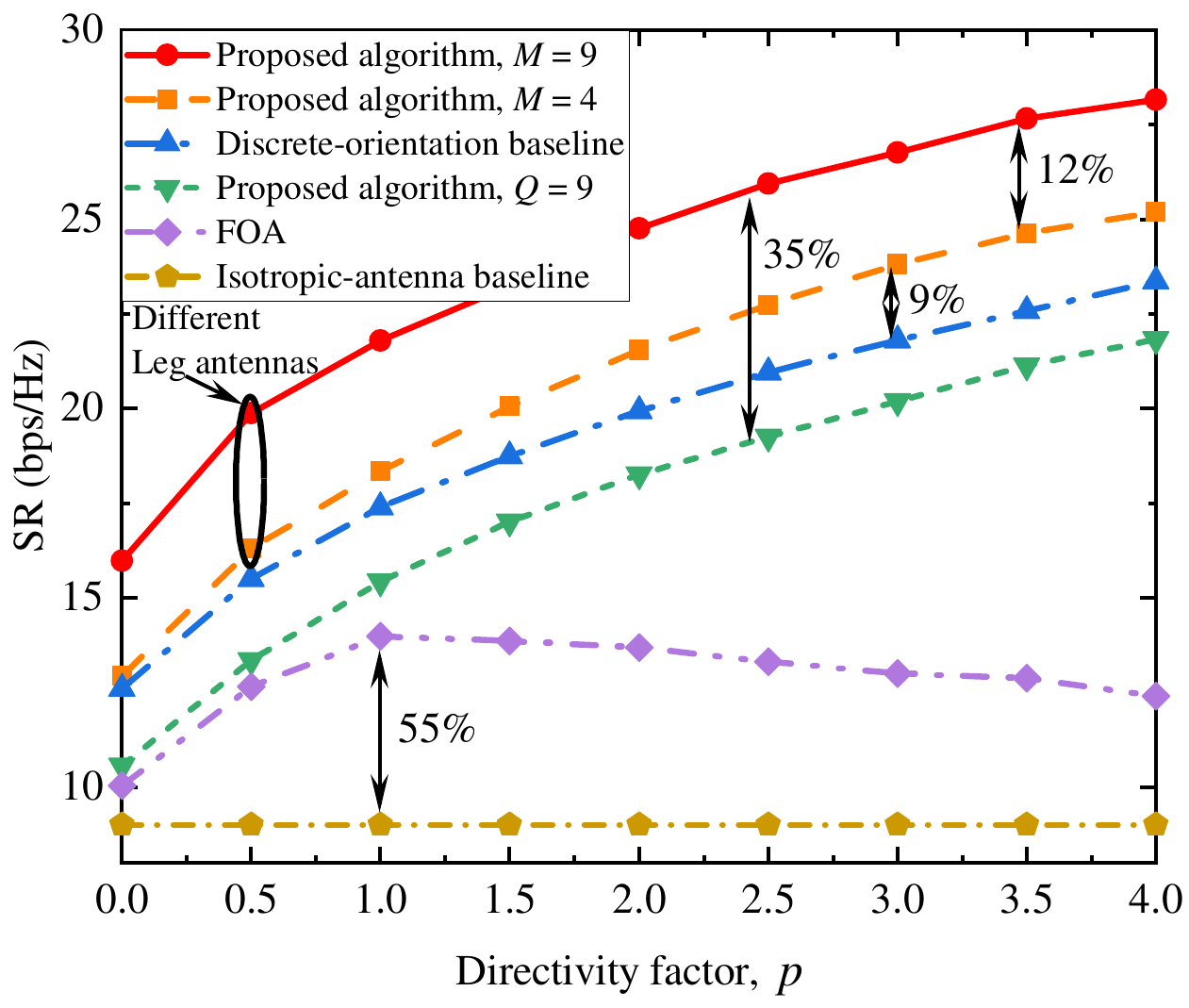}
		\caption{SR versus the directivity factor.}
		\label{Fig5}
		\vspace{-6mm}
	\end{figure}
	
	Fig.~\ref{Fig5} depicts the SR versus the antenna directivity factor \(p\) under different schemes. 
	It is observed that the SR of the directional-antenna-based schemes generally increases with \(p\). 
	Among all considered schemes, the proposed algorithm with \(M=9\) achieves the best performance, followed by the proposed algorithm with \(M=4\). 
	In contrast, the proposed algorithm with \(Q=9\) suffers from noticeable performance degradation, since more antennas at the eavesdropper strengthen its interception capability. 
	Specifically, the proposed algorithm with \(M=9\) achieves about \(35\%\) and \(12\%\) higher SR than the cases with \(Q=9\) and \(M=4\), respectively. 
	Moreover, the performance gap between the proposed algorithm and the discrete-orientation baseline gradually increases and reaches about \(9\%\) at \(p=3\).
	In addition, the isotropic-antenna scheme yields the worst performance. 
	The FOA baseline improves the SR by about \(55\%\) over the isotropic baseline at \(p=1\), but its performance decreases as \(p\) further increases. 
	This is because larger \(p\) makes the radiation energy more concentrated around the fixed boresight direction. 
	When the legitimate receiver deviates from this direction, the effective channel gain and spatial diversity are weakened. 
	Since the FOA baseline cannot adaptively steer the antenna orientations to enhance the legitimate link and suppress information leakage, its SR degrades in the high-directivity region.

	\begin{figure}[t]
		\centering
		\includegraphics[width=0.4 \textwidth]{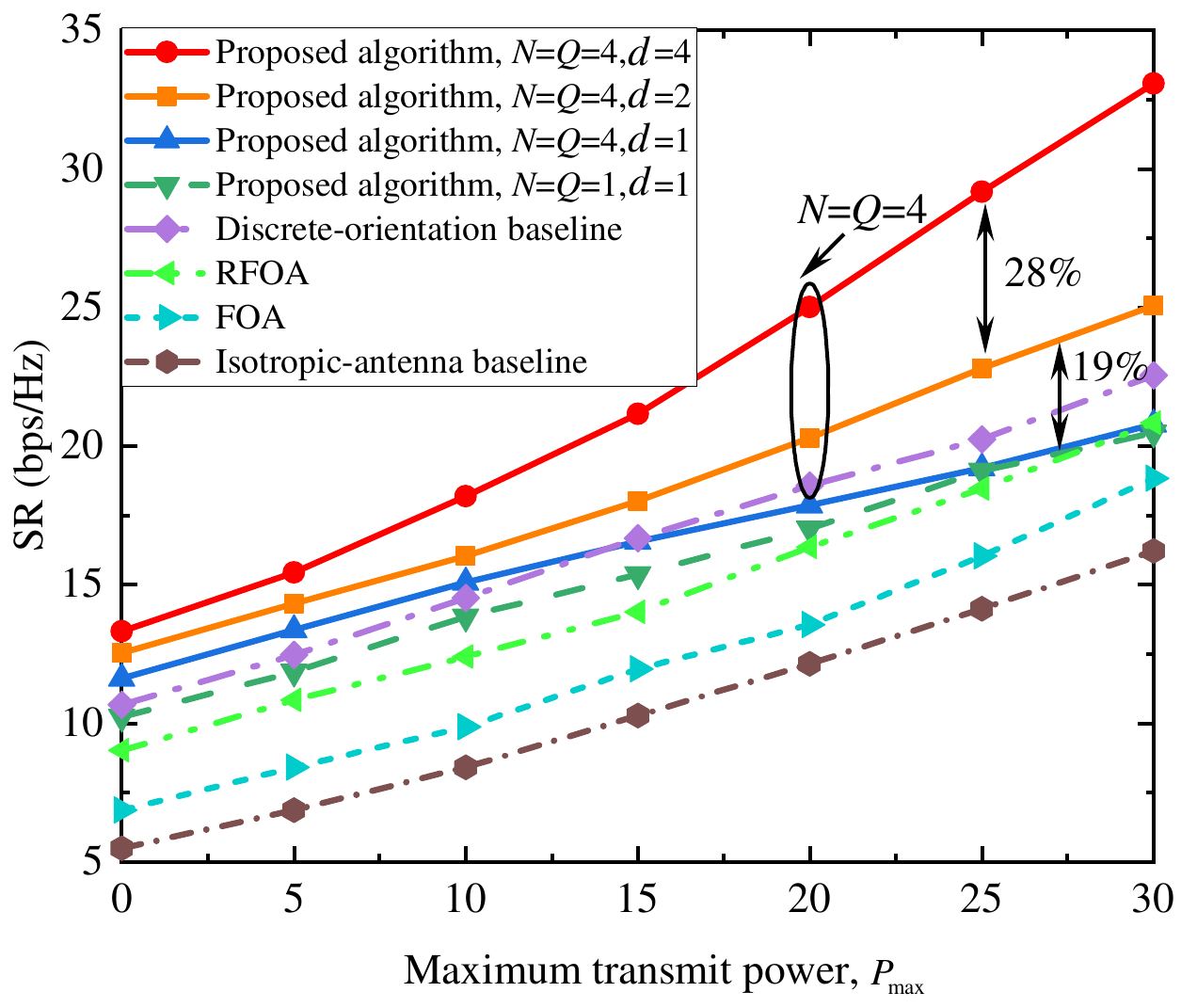}
		\caption{SR versus the maximum transmit power.}
		\label{Fig6}
		\vspace{-7mm}
	\end{figure}
	
	Fig.~\ref{Fig6} illustrates the SR versus the maximum transmit power for different schemes. 
	The SR generally increases with the maximum transmit power for all considered schemes. 
	In particular, in the low transmit-power region, increasing the number of data streams yields only limited performance improvement. By contrast, in the high transmit-power region, the performance gain brought by additional data streams becomes much more pronounced.
	This indicates that, under the same antenna configuration, increasing the number of data streams can effectively improve the SR, and the corresponding performance gain becomes more evident as the maximum transmit power increases. Specifically, when $P_{\max} = 25 ~\mathrm{dBm}$, the proposed algorithm with $N=Q=4$ and $d=4$ attains a $28\%$ higher performance gain over that with $N=Q=4$ and $d=2$, while the proposed algorithm with $N=Q=4$ and $d=1$ provides about $19\%$ higher SR than with $N=Q=1$ and $d=1$ when $P_{\max} = 30 ~\mathrm{dBm}$. 
	Furthermore, the discrete-orientation baseline with $d=2$ outperforms the proposed algorithm with $N=Q=4$ and $d=1$ when $P_{\max} \ge 15 ~\mathrm{dBm}$. This reveals that, under the same antenna configuration $N=Q=4$, increasing the number of data streams from $d=1$ to $d=2$ can provide more substantial SR improvement than merely optimizing the continuous variables with a single data stream, especially in the high transmit-power region.
	
	\begin{figure}[t]
		\centering
		\includegraphics[width=0.4 \textwidth]{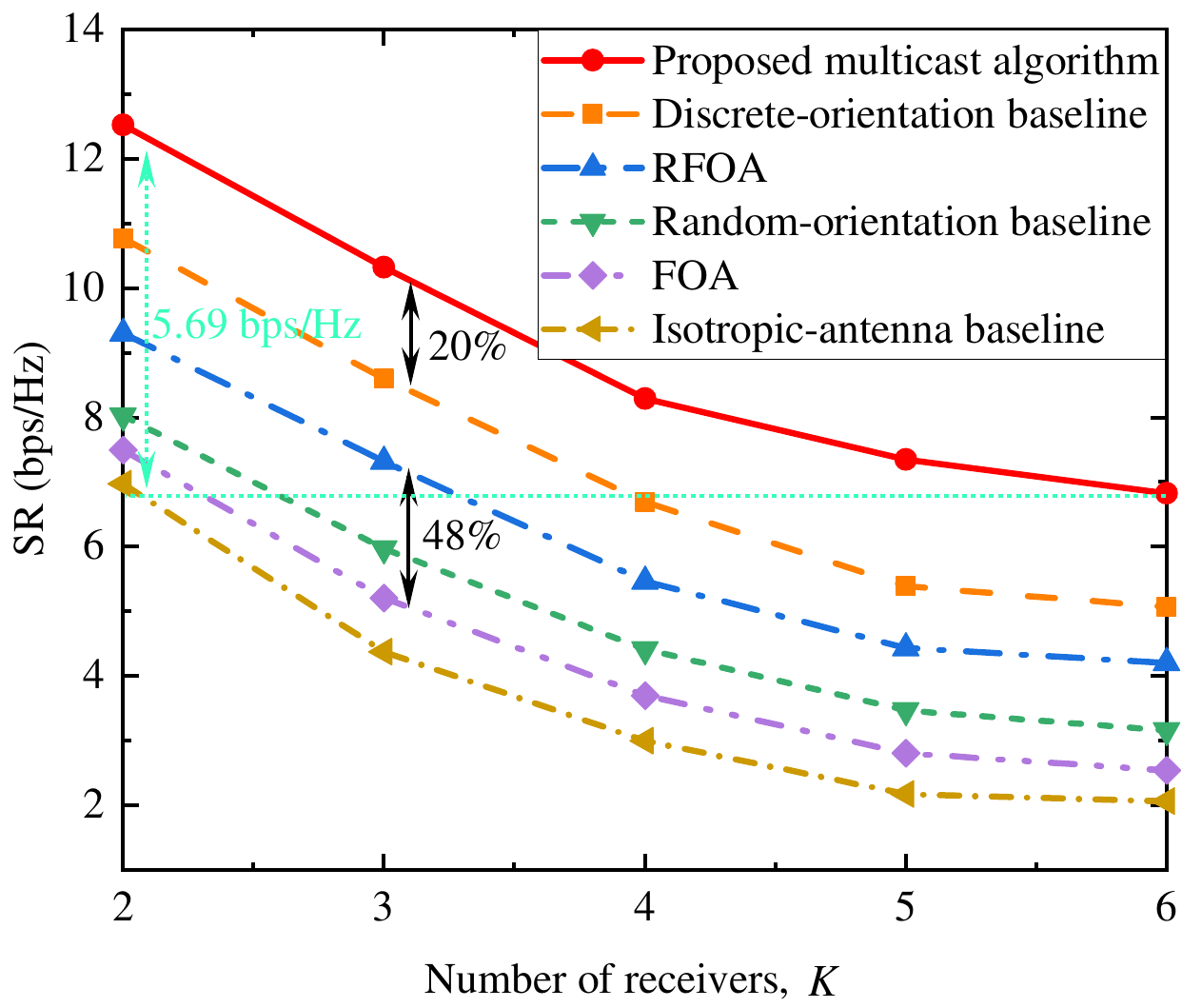}
		\caption{SR versus the number of receivers.}
		\label{Fig7}
		\vspace{-8mm}
	\end{figure}
	
	Fig. \ref{Fig7} considers the multi-receiver scenario to investigate the impact of the number of receivers on the secrecy performance of different schemes. As shown in Fig. \ref{Fig7}, the SR of all schemes decreases as the number of receivers $K$ increases. This is because, in the multicast secure transmission scenario, the achievable SR is determined by the worst-case legitimate receiver. As the number of receivers $K$ grows, it becomes more difficult to simultaneously guarantee favorable channel conditions for all receivers while suppressing information leakage, which results in a reduction in SR. When $K$ increases from $2$ to $6$, the SR of the proposed algorithm decreases by $5.69~ \mathrm{bps/Hz}$.
	Moreover, the proposed multicast algorithm always achieves the highest SR over the entire range. Specifically, the proposed multicast algorithm achieves about $20\%$ performance gain over the discrete-orientation baseline when $K=3$. Compared to the FOA baseline, the RFOA baseline provides about $48\%$ performance gain, demonstrating the benefits of the antenna rotation.
	
	\vspace{-4mm}
	\section{Conclusion}   \label{Conclusion}
	\vspace{-1mm}
	In this paper, we investigated an RA-aided secure MIMO communication system, where both the transmitter and the receiver are equipped with RAs in the presence of an eavesdropper. By exploiting the additional spatial DoFs introduced by antenna rotation, we formulated a SR maximization problem by jointly optimizing the transmit beamforming, AN covariance matrix, and transmit/receive RA orientations.
	We first studied a simplified SISO setting and revealed useful structural insights into the optimal transmit orientation. For the general MIMO case, we developed an efficient AO framework based on the MMSE reformulation, where the transmit beamforming and AN covariance matrix were updated in semi-closed forms, while the transmit and receive RA orientations were optimized on the spherical-cap manifold via the Riemannian Frank-Wolfe method. 
	We further extended the framework to the multi-receiver case by maximizing a tractable lower bound of the multicast SR.
	Simulation results demonstrated that the proposed algorithm converges rapidly and consistently outperforms conventional benchmarks.  
	
	\appendices
	\vspace{-4mm}
	\section{Determination of the Lagrange multiplier $\xi$}
	\label{Appendix A} 
	\vspace{-2mm}
	Fixed other variables,
	the beamforming update is given by
	\begin{align}
		\mathbf W(\xi)  \! =\! \left(\mathbf A \!+\! \xi \mathbf I_N\right)^{-1}\mathbf B, 
		\mathbf W_e(\xi) \!=\! \left(\mathbf A_e \!+\! \xi \mathbf I_N\right)^{-1}\mathbf B_e,
	\end{align}
	where $\mathbf A\succeq \mathbf 0$, $\mathbf A_e\succeq \mathbf 0$, and $\xi\ge 0$ is the
	Lagrange multiplier associated with the total transmit power constraint
	\begin{equation}
		\operatorname{Tr}(\mathbf W\mathbf W^H)+\operatorname{Tr}(\mathbf W_e\mathbf W_e^H) \le P_{\max}.
		\label{eq:power_constraint}
	\end{equation}
	The Lagrangian of problem $(\mathrm{\ref{P2}})$ is expressed as 
	\begin{align}
		\!\mathcal L(\mathbf W,\mathbf W_e,\xi)
		&=\operatorname{Tr}(\mathbf W^H\mathbf A\mathbf W)\!-\!2\Re\{\operatorname{Tr}(\mathbf B^H\mathbf W)\} \nonumber\\
		&+\operatorname{Tr}(\mathbf W_e^H\mathbf A_e\mathbf W_e)-2\Re\{\operatorname{Tr}(\mathbf B_e^H\mathbf W_e)\} \nonumber\\
		& +\xi\left(\operatorname{Tr}(\mathbf W\mathbf W^H)\!+\!\operatorname{Tr}(\mathbf W_e\mathbf W_e^H)\!-\!P_{\max}\right),\nonumber \\
		& \xi\ge 0.
	\end{align}
	By setting $\frac{d\mathcal L(\mathbf{W},\mathbf{W}_e)}{d \mathbf{W}^{\ast}} = 0$ and  $\frac{d\mathcal L(\mathbf{W},\mathbf{W}_e)}{d \mathbf{W}^{\ast}_e} = 0$, we have
	\begin{align}
		(\mathbf A+\xi \mathbf I_N)\mathbf W = \mathbf B,
		(\mathbf A_e+\xi \mathbf I_N)\mathbf W_e = \mathbf B_e.
	\end{align}
	The remaining KKT conditions are
	\begin{align}
		&\xi \ge 0, 
		\operatorname{Tr}(\mathbf W\mathbf W^H)+\operatorname{Tr}(\mathbf W_e\mathbf W_e^H)\le P_{\max}, \label{eq:primal_dual_feasible}\\
		&\xi\Big(\operatorname{Tr}(\mathbf W\mathbf W^H)+\operatorname{Tr}(\mathbf W_e\mathbf W_e^H)-P_{\max}\Big)=0.
		\label{eq:comp_slackness}
	\end{align}
	Substitute $\mathbf W(\xi)$ and $\mathbf W_e(\xi)$ into \eqref{eq:primal_dual_feasible}, we have 
	\begin{align}
		P(\xi)
		&\triangleq \operatorname{Tr}\!\big(\mathbf W(\xi)\mathbf W(\xi)^H\big)
		+\operatorname{Tr}\!\big(\mathbf W_e(\xi)\mathbf W_e(\xi)^H\big) \nonumber\\
		&= \operatorname{Tr}\!\Big((\mathbf A+\xi\mathbf I_N)^{-1}\mathbf B\mathbf B^H(\mathbf A+\xi\mathbf I_N)^{-1}\Big) \nonumber\\
		&+\operatorname{Tr}\!\Big((\mathbf A_e+\xi\mathbf I_N)^{-1}\mathbf B_e\mathbf B_e^H(\mathbf A_e+\xi\mathbf I_N)^{-1}\Big).
		\label{eq:Plambda}
	\end{align}
	By setting $\xi=0$, the unconstrained solution is given by
	\begin{equation}
		\mathbf W(0)=\mathbf A^\dagger \mathbf B, \mathbf W_e(0)=\mathbf A_e^\dagger \mathbf B_e.
	\end{equation}
	If	$P(0)\le P_{\max}$, the power constraint is inactive and the optimal multiplier is $ \xi^\star=0$.

	When $P(0)>P_{\max}$, the constraint must be active at optimum, we have $P(\xi^\star)=P_{\max}$ and $ \xi^\star>0$.
	For $\mathbf A\succeq 0$, define $\mathbf X(\xi)=(\mathbf A+\xi \mathbf I_N)^{-1}$, $\xi>0$.
	Then $\mathbf X(\xi)$ is hermitian positive semidefinite (PSD) and strictly decreases in the lower order with $\xi$.
	Consequently, for any PSD matrix $\mathbf C\succeq 0$, $
	g(\xi)=\operatorname{Tr}\!\big(\mathbf X(\xi)\mathbf C\mathbf X(\xi)\big)$
	is non-increasing in $\xi$. Since $\mathbf B\mathbf B^H\succeq 0$ and $\mathbf B_e\mathbf B_e^H\succeq 0$,
	both trace terms in \eqref{eq:Plambda} are non-increasing, hence $P(\xi)$ is continuous and strictly
	decreasing for $\xi>0$ (unless $\mathbf B$ or $\mathbf B_e$ is zero). Therefore, $P(\xi^\star)=P_{\max}$
	has a unique solution.
	
	Since $\mathbf A+\xi\mathbf I_N \succeq \xi\mathbf I_N$ for $\xi>0$, we have
	\begin{equation}
		(\mathbf A+\xi\mathbf I_N)^{-1} \preceq \frac{1}{\xi}\mathbf I_N,
		(\mathbf A_e+\xi\mathbf I_N)^{-1} \preceq \frac{1}{\xi}\mathbf I_N.
	\end{equation}
	Then, we get
	\begin{align}
		&\operatorname{Tr}\!\Big((\mathbf A+\xi\mathbf I_N)^{-1}\mathbf B\mathbf B^H(\mathbf A+\xi\mathbf I_N)^{-1}\Big) \nonumber \\
		&\le \operatorname{Tr}\!\Big(\frac{1}{\xi}\mathbf I_N\ \mathbf B\mathbf B^H\ \frac{1}{\xi}\mathbf I_N\Big)
		= \frac{1}{\xi^2}\operatorname{Tr}(\mathbf B\mathbf B^H),\\
		&\operatorname{Tr}\!\Big((\mathbf A_e+\xi\mathbf I_N)^{-1}\mathbf B_e\mathbf B_e^H(\mathbf A_e+\xi\mathbf I_N)^{-1}\Big)
		\le \frac{1}{\xi^2}\operatorname{Tr}(\mathbf B_e\mathbf B_e^H). \nonumber
	\end{align}
	Therefore, we have
	\begin{equation}
		P(\xi)\le \frac{1}{\xi^2}\Big(\operatorname{Tr}(\mathbf B\mathbf B^H)+\operatorname{Tr}(\mathbf B_e\mathbf B_e^H)\Big).
	\end{equation}
	The upper bound of $\xi$ is given by
	\begin{equation}
		\xi_{\mathrm{ub}}
		\ \triangleq\
		\sqrt{\frac{\operatorname{Tr}(\mathbf B\mathbf B^H)+\operatorname{Tr}(\mathbf B_e\mathbf B_e^H)}{P_{\max}}}.
		\label{eq:lambda_ub}
	\end{equation}
	Then $P(\xi_{\mathrm{ub}})\le P_{\max}$ is guaranteed, so $\xi^\star \in (0,\xi_{\mathrm{ub}}]$.
	The bisection method can be applied to find the best $\xi^{\star}$.
	
	\vspace{-6mm}
	\section{Channel Derivatives} \label{Appendix B}
	\vspace{-2mm}
	For the LoS component $h_{m,n}^{\text{LoS}}$, we have 
	\begin{align}
		\label{h_k_gradient}
		 &\frac{\partial h_{m,n}^{\text{LoS}}}{\partial \mathbf{f}_{t,n}} \!\!=\!\! a_{m,n} f_{r,m,n} \frac{\partial f_{t,m,n}}{\partial \mathbf{f}_{t,n}},  \frac{\partial h_{m,n}^{\text{LoS}}}{\partial \mathbf{f}_{r,m}} \!\!=\!\! a_{m,n} f_{t,m,n} \frac{\partial f_{r,m,n}}{\partial \mathbf{f}_{r,m}}, \nonumber \\
		&\frac{\partial f_{t,m,n}}{\partial \mathbf{f}_{t,n}} \!\!=\!\!
		\begin{cases}
			p \left[\tilde{\mathbf{f}}_{t,n}^\mathrm{T}\mathbf{d}_{m,n} \right]^{p-1} \!\!\!\!\! \mathbf{R}_t^\mathrm{T}\mathbf{d}_{m,n}, \!\!&\!\! \text{if } \tilde{\mathbf{f}}_{t,n}^\mathrm{T}\mathbf{d}_{m,n} \!>\! 0, \\
			0, \!\!&\!\! \text{otherwise},
		\end{cases} \nonumber \\
		&\frac{\partial f_{r,m,n}}{\partial \mathbf{f}_{r,m}} \!\!=\!\!
		\begin{cases}
			\!-p \left[ \!-\tilde{\mathbf{f}}_{r,m}^\mathrm{T}\mathbf{d}_{m,n} \right]^{p-1} \!\!\! \mathbf{R}_{r}^\mathrm{T}\mathbf{d}_{m,n}, \!\!&\!\!\! \text{if } \!-\!\tilde{\mathbf{f}}_{r,m}^\mathrm{T}\mathbf{d}_{m,n} \!>\! 0, \\
			0, \!\!&\!\! \text{otherwise},
		\end{cases} \nonumber
	\end{align}
	where $\tilde{\mathbf{f}}_{t,n} = \mathbf{R}_t\mathbf{f}_{t,n}$, $\tilde{\mathbf{f}}_{r,m} = \mathbf{R}_{r}\mathbf{f}_{r,m}$, and  $\mathbf{d}_{m,n} = \frac{\mathbf{r}_{m}-\mathbf{t}_n}{r_{m,n}}$.
	
	For the NLoS component $h_{k,m,n}^{\text{NLoS}}$, we have
	\begin{align}
		\frac{\partial h_{m,n}^{\text{NLoS}}}{\partial \mathbf{f}_{t,n}} &= \sum\nolimits_{d=1}^D b_{m,n,d} f_{r,d,m} \frac{\partial f_{t,n,d}}{\partial \mathbf{f}_{t,n}}, \\
		\frac{\partial f_{t,n,d}}{\partial \mathbf{f}_{t,n}} &\!\!=\!\!
		\begin{cases}
			p \left[ \tilde{\mathbf{f}}_{t,n}^\mathrm{T}\mathbf{d}_{n,d} \right]^{p-1} \!\!\!\!\! \mathbf{R}_t^\mathrm{T}\mathbf{d}_{n,d}, & \text{if } \tilde{\mathbf{f}}_{t,n}^\mathrm{T}\mathbf{d}_{n,d} \!>\! 0, \\
			0, & \text{otherwise},
		\end{cases}\\
		\label{h_k_NLOS-gradient}
		\frac{\partial h_{m,n}^{\text{NLoS}}}{\partial \mathbf{f}_{r,m}} &= \sum\nolimits_{d=1}^D b_{m,n,d} f_{t,n,d} \frac{\partial f_{r,d,m}}{\partial \mathbf{f}_{r,m}}, \\
		\frac{\partial f_{r,d,m}}{\partial \mathbf{f}_{r,m}} &\!\!=\!\!
		\begin{cases}
			p \left[\tilde{\mathbf{f}}_{r,m}^\mathrm{T}\mathbf{d}_{m,d} \right]^{p-1}\!\!\!\! \mathbf{R}_{r}^\mathrm{T}\mathbf{d}_{m,d}, \!\!&\!\!\! \text{if } \tilde{\mathbf{f}}_{r,m}^\mathrm{T}\mathbf{d}_{m,d} \!>\! 0, \\
			0, \!\!&\!\!\! \text{otherwise},
		\end{cases}
	\end{align}
	where $\mathbf{d}_{n,d} =\frac{\mathbf{s}_d-\mathbf{t}_n}{r_{n,d}}$ and $\mathbf{d}_{m,d} = \frac{\mathbf{s}_d-\mathbf{r}_{m}}{r_{d,m}}$.
	
	For eavesdropper channels $h_{e,q,n}$, let $\mathbf{d}_{e,q,n} =\frac{\mathbf{e}_{q}-\mathbf{t}_n}{r_{e,q,n}}$, the LoS and NLoS components are given by
	\begin{align}
		\!\!\!\!\frac{\partial h_{e,q,n}^{\text{LoS}}}{\partial \mathbf{f}_{t,n}} &\!\!=\!\! a_{e,q,n}\frac{\partial f_{e,q,n}}{\partial \mathbf{f}_{t,n}}, \frac{\partial h_{e,q,n}^{\text{NLoS}}}{\partial \mathbf{f}_{t,n}} \!\! =\!\! \sum\nolimits_{d=1}^D \!\!\! b_{e,q,n,d}\frac{\partial f_{t,n,d}}{\partial \mathbf{f}_{t,n}}, \\
		\!\!\!\!\frac{\partial f_{e,q,n}}{\partial \mathbf{f}_{t,n}} &\!\!=\!\!
		\begin{cases}
			p \left[ \tilde{\mathbf{f}}_{t,n}^\mathrm{T}\mathbf{d}_{e,q,n} \right]^{p-1}\!\!\!\! \mathbf{R}_t^\mathrm{T}\mathbf{d}_{e,q,n}, \!\!&\!\! \text{if } \tilde{\mathbf{f}}_{t,n}^\mathrm{T}\mathbf{d}_{e,q,n} \!>\! 0, \\
			0, \!\!&\!\! \text{otherwise}.
		\end{cases}
	\end{align}
	
	\vspace{-2mm}
	\bibliographystyle{IEEEtran}
	\vspace{-2mm}
	\begin{spacing}{0.89}
		\bibliography{thesis}
	\end{spacing}
	
\end{document}